\documentclass[prb,twocolumn]{revtex4-1}

\usepackage{mathtools}
\usepackage{amsmath,bm}
\usepackage{amssymb}
\usepackage{subfigure}
\usepackage{setspace}
\usepackage{dcolumn}
\usepackage{color}

\usepackage[hidelinks]{hyperref}

\begin{document}

\title{Adiabatic pumping via avoided crossings in stiffness modulated quasiperiodic beams}

\author{Emanuele Riva$^{a}$}
\email[]{emanuele.riva@polimi.it}
\affiliation{ $^a$ Department of Mechanical Engineering, Politecnico di Milano, Italy, 20156}
\author{Vito Casieri$^{a}$}
\affiliation{ $^a$ Department of Mechanical Engineering, Politecnico di Milano, Italy, 20156}
\author{Ferruccio Resta$^{a}$}
\affiliation{ $^a$ Department of Mechanical Engineering, Politecnico di Milano, Italy, 20156}
\author{Francesco Braghin$^{a}$}
\affiliation{ $^a$ Department of Mechanical Engineering, Politecnico di Milano, Italy, 20156}


\date{\today}

\begin{abstract}
In this manuscript we report on adiabatic pumping in quasiperiodic stiffness modulated beams. We show that distinct topological states populating nontrivial gaps can nucleate \textit{avoided crossings} characterized by edge-to-edge transitions. Such states are inherently coupled when a smooth variation of the modulation phase is induced along a synthetic dimension, resulting in topological edge-to-edge transport stemming from distinct polarizations of the crossing states. 
We first present a general framework to estimate the required modulation speed for a given transition probability in time. Then, this analysis tool is exploited to tailor topological pumping in a stiffness modulated beam. 
\end{abstract}

\maketitle

\section{Introduction}\label{Introduction}
The study of topological insulators in physics has gained great importance in the past years, due to the opportunity to achieve defect immune and lossless energy transport within different research fields and physical platforms, such as photonics \cite{lu2014topological,khanikaev2013photonic}, quantum systems \cite{hasan2010colloquium,haldane1988model}, and acoustics \cite{PhysRevLett.114.114301,fleury2016floquet,lu2017observation,mousavi2015topologically} among others.
In mechanics, topologically protected edge waves have been extensively studied in analogy with quantum systems. Indeed, the systematic combination of topology to the study of nontrivial band structures has opened a new branch of studies under the name of topological mechanics \cite{huber2016topological}. Notable examples include elastic analogues to the \textit{Quantum Hall Effect} (QHE) \cite{wang2015topological,nash2015topological,souslov2017topological,chen2019mechanical}, the \textit{Quantum Spin Hall Effect} (QSHE) \cite{susstrunk2015observation,pal2016helical,PhysRevB.98.094302,PhysRevX.8.031074} and \textit{Quantum Valley Hall Effect} (QVHE) \cite{pal2017edge,vila2017observation,liu2018tunable,liu2019experimental,riva2018tunable}, which are associated to robust propagation mechanisms of technological relevance for next generation applications involving elastic wave manipulation, isolation and waveguiding.\\
Other approaches to topology-based design leverage nontrivial topological properties emerging from a relevant higher-order parameter space \cite{kraus2016quasiperiodicity,ozawa2016synthetic,lee2018electromagnetic}. 
In this context, the projection of a nontrivial topology to a physical set of parameters reflects on modulation families (either spatial or spatiotemporal), which can be exploited to manipulate wave propagation \cite{kraus2012topological,zilberberg2018photonic}. 
That is, according to the bulk-edge corresponding principle, the formation of localized edge states is inherently linked with the topological characteristics of the wavenumber-parameter space \cite{apigo2019observation,ni2019observation,apigo2018topological}. In other words, the edge state localization is parameterized through a projection phase \cite{Pal_2019,xia2020topological}. When such parameter is smoothly varied along a second dimension, the edge state transforms from being left (right) to right (left) localized, therefore establishing a topological pump \cite{brouzos2019non,verbin2015topological,nakajima2016topological,lohse2016thouless,longhi2019topological}.
Recent examples include mechanical lattices with periodic couplings \cite{rosa2019edge}, elastic plates with smoothly varying square-wave modulations \cite{PhysRevB.101.094307}, and magneto-mechanical structures with time-varying parameters \cite{grinberg2020robust}. 
In general, the adiabatic transformation of the edge state is necessary for a successful realization of a topological waveguide, which require a slow variation of the phase parameter in space or time. 
In the attempt to provide an estimate of the required speed of modulation in edge-to-edge transformations, we study a quasiperiodic stiffness modulated beam. This specific configuration supports topological boundary modes, whose frequency and mode polarization is function of a modulation phase parameter, which is suitably varied in time through established techniques \cite{riva2019generalized,marconi2019experimental,riva2020non,nassar2018quantization}. It is illustrated that distinct topological modes can populate the same gap and - through a smooth variation of the phase parameter in time - can nucleate crossing states, also known as \textit{avoided crossing}, in which the corresponding mode polarizations couple with each other, expanding the range of opportunities in topology-based waveguiding.
Moreover, the required speed of modulation for a given transition probability is estimated as a function of few critical parameters of the crossing states. We demonstrate that, depending upon the phase speed of variation along the temporal dimension, a localized state can simply cross the intesection (fast modulation) without shape modification, or can split in two separate states localized at both edges (intermediate speed) or can fully transform into a state localized at the opposite boundary (slow modulation). To this end, we first present the theoretical framework to compute transition probabilities applied to a simple spring mass system through a paraxial approximation of the equation of motion. 
Then, the same theory is applied to study adiabatic and non-adiabatic transformations in the quasiperiodic beam.\\
This study is relevant for the optimization of pumping protocols in mechanics, which suit applications involving wave splitting and de-multiplexing, such as nondestructive evaluation, signal transmission and realization of logic circuits based on elastic wave propagation.

\section{Adiabatic transformations through avoided crossings}
\label{section2}
\begin{figure*}[t!]
	\centering
	\hspace{-0.4cm}
	\subfigure[]{\includegraphics[width=0.34\textwidth]{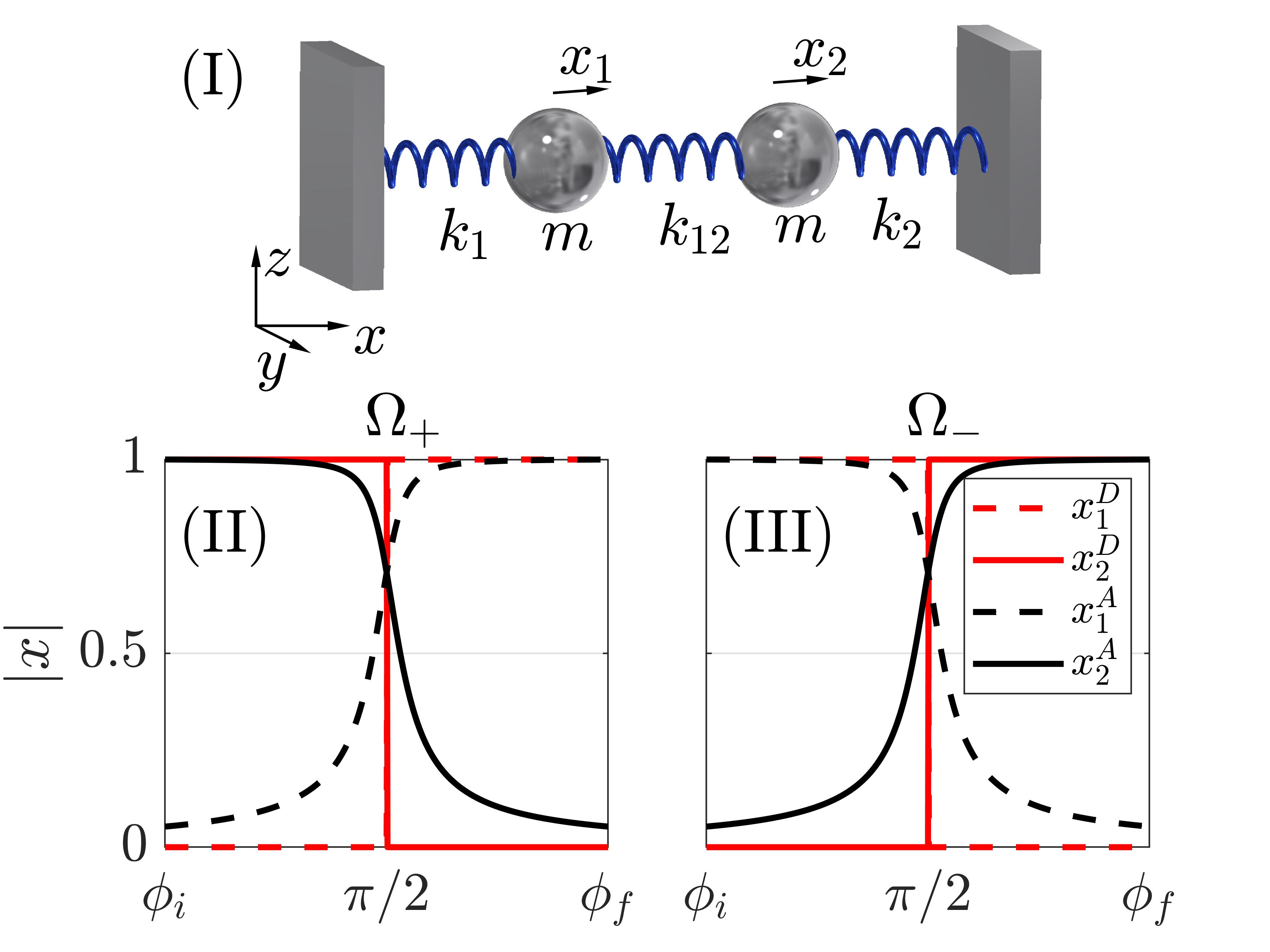}\label{Fig1a}}
	\hspace{-0.4cm}
	\subfigure[]{\includegraphics[width=0.34\textwidth]{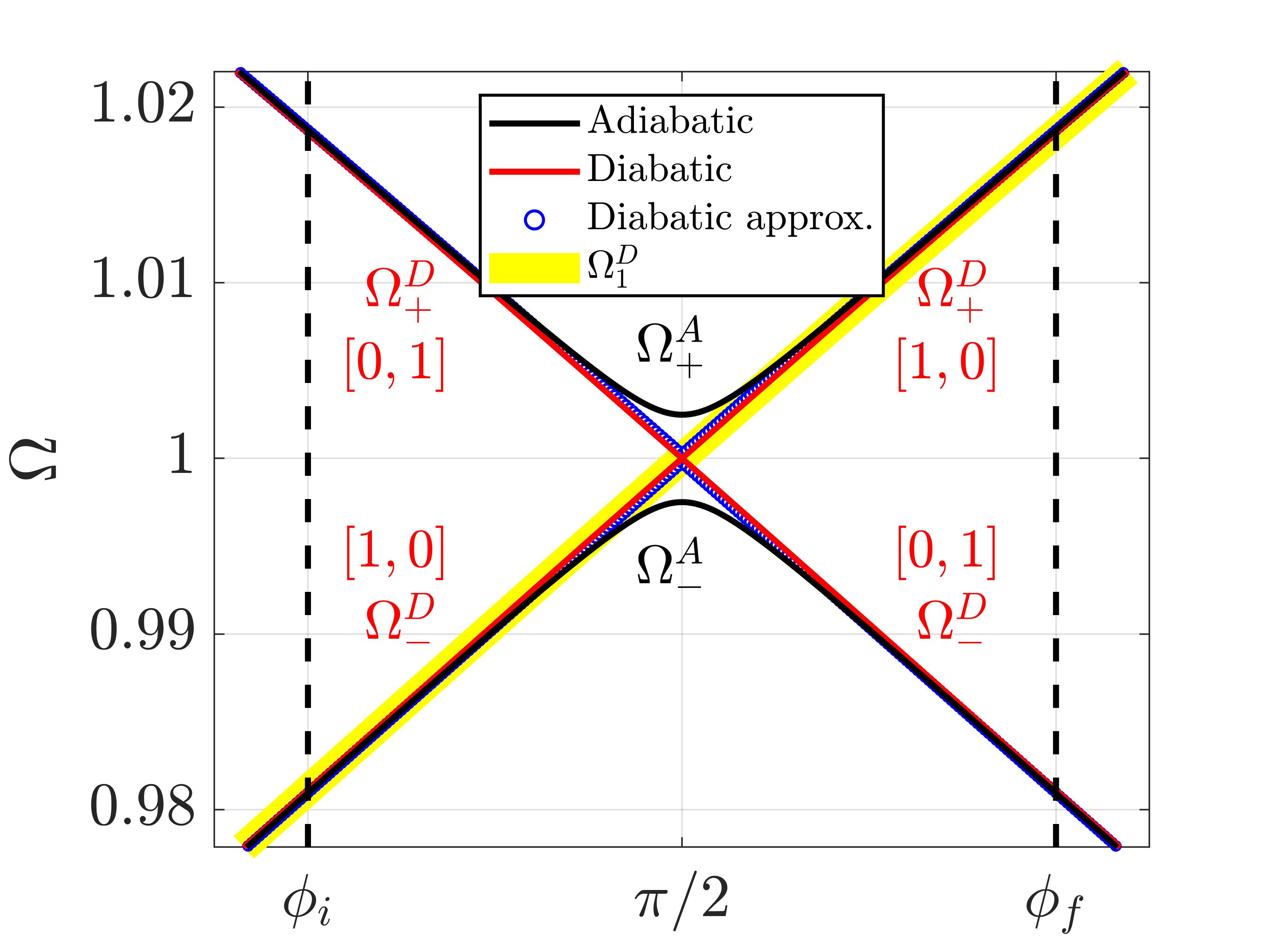}\label{Fig1a}}
	\hspace{-0.4cm}	\subfigure[]{\includegraphics[width=0.34\textwidth]{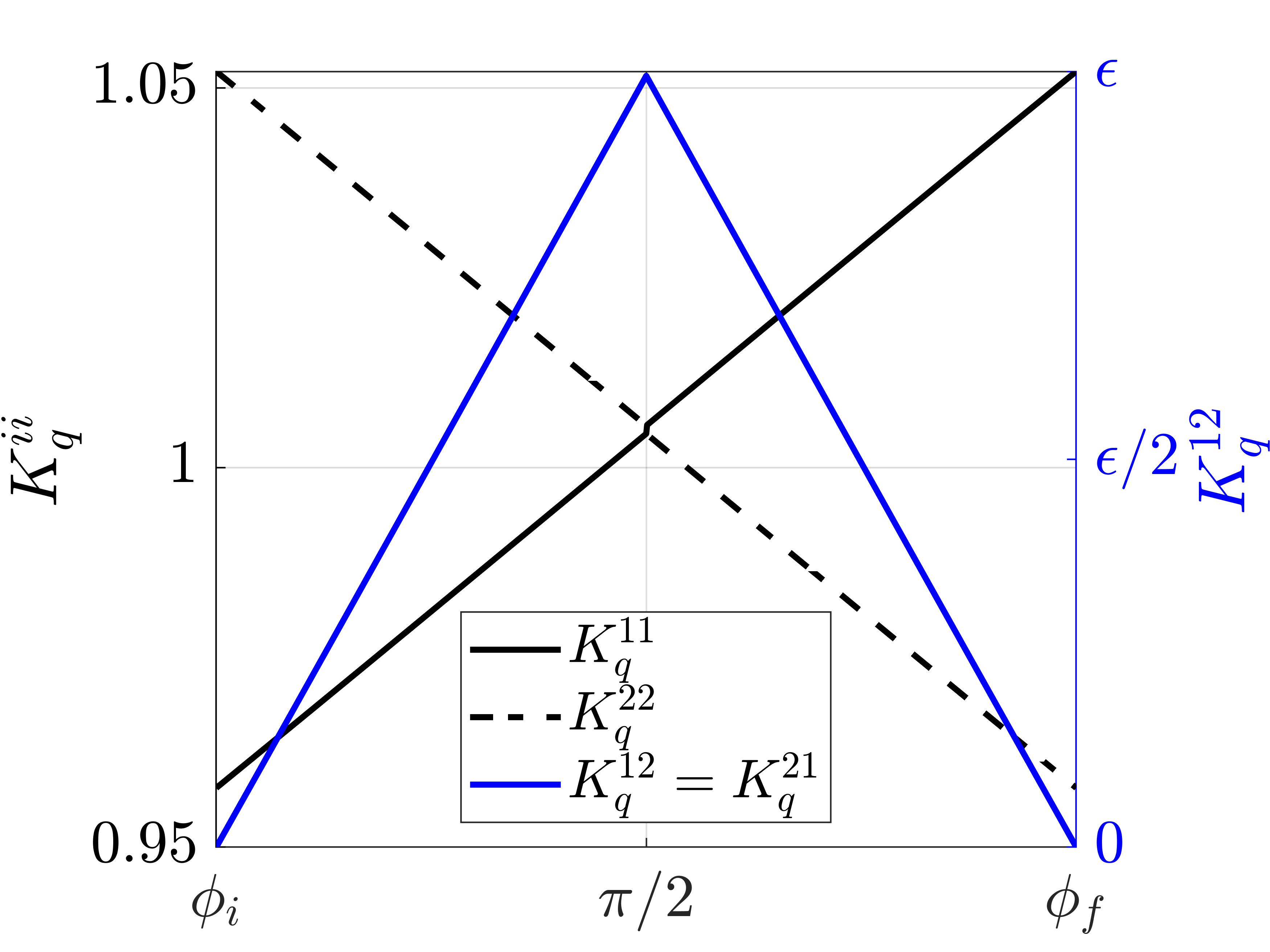}\label{Fig1a}}
	\caption{(a)-${\rm I}$ Schematic of the spring-mass system. The adopted parameters are: $m=1$, $\phi_i=0.45\pi$, $\phi_f=0.55\pi$, $k_0=1$, $\alpha=0.3$ $\varepsilon=0.005$ (a)-${\rm II}$ Diabatic and adiabatic mode polarizations ${\bm x}^D=\left[x_1^D,x_2^D\right]^T$ and ${\bm x}^A=\left[x_1^A,x_2^A\right]^T$ associated to the top branch $\Omega_+$ (a)-${\rm III}$ and to the bottom branch $\Omega_-$. Displacements relative to diabatic states are represented with red lines, while adiabatic states are illustrated with black curves. (b) Adiabatic, diabatic and approximated diabatic states upon varying the modulation phase parameter $\phi\in\left[\phi_i,\phi_f\right]$. The yellow band represents the locus of diabatic states $\Omega_1^D$ characterized by the same polarization ${\bm{x}}^D\approx[1,0]$.  (c) Estimated coefficients of the stiffness matrix in generalized coordinates. $\hat{K}_{1,1}$ and $\hat{K}_{2,2}$ are the diagonal terms, which are displayed with solid and dashed black lines, respectively. $\hat{K}_{1,2}=\hat{K}_{2,1}$ are the off-diagonal terms (blue line) representing the coupling between states.}
	\label{Fig1}
\end{figure*}
We start considering a simple 2 mass-spring system illustrated in Fig. \ref{Fig1}(a)-${\rm I}$ in which the point masses $m_1$ and $m_2$ are respectively grounded through linear springs $k_1\left(\phi(t)\right)=k_0\left[1-\alpha\cos{\left(\phi\right)}\right]$ and $k_2\left(\phi(t)\right)=k_0\left[1+\alpha\cos{\left(\phi\right)}\right]$ that are smooth functions of time through a control phase parameter $\phi\left(t\right)=\phi_i+\omega_m t$, being $\phi_i$ and $\omega_m$ the initial phase and the angular velocity, respectively. $\alpha$ is the stiffness modulation amplitude. It is assumed that $m_1=m_2=m$. In addition, a linear time-invariant spring $k_{12}=\varepsilon k_0$ is placed between the first and second mass and represents a weak coupling between the mass displacements $x_1$ and $x_2$, for a sufficiently small value of $\varepsilon$.
Upon linear momentum balance, one can write the elasto-dynamic equations governing the motion of the system:
\begin{equation}
\begin{split}
&\ddot{x}_1+\omega_1^2x_1-\Gamma\sqrt{\omega_1\omega_2}x_2=0\\
&\ddot{x}_2-\Gamma\sqrt{\omega_1\omega_2}x_1+\omega_2^2x_2=0
\end{split}
\label{eq_01}
\end{equation}
in which $\omega_{1,2}=\sqrt{\left(k_{1,2}+k_{12}\right)/m}$ and the off-diagonal terms are independent on time. $\Gamma=k_{12}/\displaystyle\sqrt{m^2\omega_1\omega_2}$ represents the coupling coefficient, which is analogue to Rabi's frequency for quantum systems. Eq. \ref{eq_01} is written in compact form:
\begin{equation}
\begin{split}
{\bm M}{\ddot{\bm x}} + {\bm K}{\bm x}=0
\end{split}
\label{eq_01_1}
\end{equation}
one can seek Ansatz solutions in the form ${\bm x}={\hat{\bm x}}_0{\rm e}^{{\rm i}\omega t}$ yielding the \textit{adiabatic frequencies} $\omega_{\pm}$, i.e. the frequencies corresponding to the coupled states through the parameter $\Gamma$. That is:
\begin{equation}
\omega_{\pm}=\sqrt{\frac{1}{2}\left[\omega_1^2+\omega_2^2\pm\sqrt{\left(\omega_1^2-\omega_2^2\right)^2+4\Gamma^2\omega_1\omega_2}\right]}
\label{eq_02}
\end{equation}
it is evident that, for $\Gamma=0$, one gets $\omega_{\pm}^D=\omega_{1,2}$, leading to uncoupled states. Under this condition, $\omega_{\pm}^D$ are known as \textit{diabatic frequencies}. 
The location of the states is mapped through Eq. \ref{eq_02} upon varying the modulation phase $\phi$ and is illustrated in Fig \ref{Fig1}(b), where $\Omega^D\left(\phi\right)_{\pm}=\omega_{\pm}^D\left(\phi\right)/\omega_0$ is a dimensionless frequency, with $\omega_0=\sqrt{k_0/m}$. Similarly, a dimensionless modulation speed $\Omega_m=\omega_m/\omega_0$ is defined.
The investigation domain is limited in the neighborhood of $\phi=\pi/2$ around which, for $\Gamma=0$, the lower and upper diabatic frequencies $\Omega_-^D$ and $\Omega_+^D$ (represented in red) are linear and coincident for $\Omega_\pm^D=1$.
Interestingly, for $\phi=\pi/2$ and $\Omega_\pm^D=1$ the spectrum undergoes a transformation such that the associated mode polarizations ${\bm x}^D_-\left(\phi_i\right)=\left[1,0\right]^T$ and ${\bm x}^D_+\left(\phi_i\right)=[0,1]^T$ interchange each other, i.e. ${\bm x}^D_-\left(\phi_i\right)\rightarrow{\bm x}^D_+\left(\phi_f\right)$ and ${\bm x}^D_+\left(\phi_i\right)\rightarrow{\bm x}^D_-\left(\phi_f\right)$. To elucidate this concept, ${\bm x}^D_-$ and ${\bm x}^D_+$ are displayed in Figs. \ref{Fig1}(a) ${\rm II-III}$ employing dashed and solid red curves for $x_1^D$ and $x_2^D$, respectively.\\
In contrast, $\Gamma\neq 0$ implies opening of the crossing cone and coupling between otherwise degenerate states.  The adiabatic spectrum $\Omega_{\pm}^A\left(\phi\right)$ emerges when nonzero coupling is considered and, specifically, it deviates from the diabatic spectrum when the coupling between states is stronger, as shown by the black curve in Fig. \ref{Fig1}(b). The corresponding mode polarizations $\bm{x}^A_-$ and $\bm{x}^A_+$, represented by the black curves in Fig. \ref{Fig1}(a) ${\rm II-III}$, can be regarded as smooth perturbation to the diabatic solutions due to weak coupling $\Gamma$.
In other words, one can evaluate diabatic states by assuming adiabatic solutions and nullifying $\Gamma$, which is generally unknown in more complicated case-studies, such as quasiperiodic systems. 
To overcome this issue, we hereafter present a systematic approximation of diabatic states and coupling parameter, which will be used later in the paper to estimate transition probabilities in adiabatic transformations. Let us consider the system for $\phi=\phi_i$, which is sufficiently far away from $\phi=\pi/2$, such that the adiabatic and diabatic states $\Omega_{\pm}^A\left(\phi_i\right)$ and $\Omega_{\pm}^D\left(\phi_i\right)$ and corresponding polarizations ${\bm x}^D_{\pm}\left(\phi_i\right)$ and ${\bm x}^A_{\pm}\left(\phi_i\right)$ are approximately coincident. We also observe that, if $\Gamma=0$, the \textit{diabatic} eigenvector basis  ${\bm \Psi}^D=[{\bm x}^D_{-}\left(\phi\right),{\bm x}^D_{+}\left(\phi\right)]$ preserves unaltered with $\phi$ (except from the crossing point at $\phi=\pi/2$, in which the eigenvectors simply exchanges each others). This implies that, under of change of coordinates ${\bm x}={\bm \Psi}^D{\bm q}$, the modal displacements $\bm q$ are uncoupled for any $\phi$ value, that is:
\begin{equation}
\left(-\omega^2{\bm M}_q+{\bm K}_q\left(\phi\right)\right){\bm q}=0
\label{eq_04}
\end{equation}
being ${\bm M}_q=({\bm \Psi}^D)^T{\bm I}{\bm \Psi}^D$ and ${\bm K}_q\left(\phi\right)=({\bm\Psi}^D)^T{\bm K}{\bm \Psi}^D$ diagonal mass and stiffness matrices, respectively. 
In contrast, if $\Gamma\neq0$ the \textit{adiabatic} eigenvector basis ${\bm \Psi}^A$ undergoes a smooth modification for $\phi_i\rightarrow\phi_f$ (see Fig. \ref{Fig1}(a) ${\rm II-III}$), starting from initial and final values that well approximate the diabatic basis ${\bm \Psi}^D$.\\
Enforcing ${\bm \Psi}^D_{i,f}\approx{\bm \Psi}^A_{i,f}$, a new change of coordinates ${\bm x}={\bm \Psi}^A_{i,f}{\bm q}$ reflects into a symmetric stiffness matrix ${\bm K}_q\left(\phi\right)=({\bm \Psi}^A)^T{\bm I}{\bm\Psi}^A$, in which the off-diagonal terms embody the modal coupling $K_{q}^{12}=K_{q}^{21}$. 
The coefficients of ${\bm K}_q\left(\phi\right)$ are illustrated in Fig. \ref{Fig1}(c) and are evaluated employing the change of coordinates ${\bm x}={\bm\Psi}_i^A{\bm q}$ for $\phi_i<\phi<\pi/2$ and ${\bm x}={\bm\Psi}_f^A{\bm q}$ for $\pi/2<\phi<\phi_f$ to compensate for the eigenfrequency interchange. As expected, $K_{q}^{12}$ reaches the maximum value of $\epsilon$ for $\phi=\pi/2$ and is responsible for the frequency and shape difference between diabatic and adiabatic states. 
It is therefore straightforward to conclude that enforcing $K_{q}^{12}=K_{q}^{21}=0$ into Eq. \ref{eq_04}, the modal coupling breaks and, as a result, one gets approximated diabatic frequencies which, in turn, are illustrated with blue dots in Fig \ref{Fig1}(b).
It is worth mentioning that this procedure yields an estimation of the coupling, as ${\bm\Psi}^A_{i,f}$ only approximates ${\bm \Psi}^D_{i,f}$. Such estimation becomes more accurate as the adiabatic basis ${\bm\Psi}^A_{i,f}$ converges to ${\bm\Psi}^D_{i,f}$.\\
Now, in the attempt to find the transition probabilities for an eigensolution $\Omega_-^A\left(\phi_i\right)$ for $t=0$ belonging to the bottom branch $\Omega_-^A$ to jump to the upper branch $\Omega_+^A$ for smooth modulations $\phi\left(t\right)=\phi_i+\omega_m t$, we proceed with the following approximation of the equation of motion, with the aim to present a mechanical analogue to the Landau-Zener model. We remark that, at this step, the temporal evolution of the diabatic states $\Omega_{1,2}^{D}\left(t\right)$ is known and corresponds to the path that preserves the starting mode polarization unaltered through ${\bm x}^{D}_i\rightarrow{\bm x}^{D}_f$, which is highlighted with a yellow band for $\Omega_1^D\left(t\right)$ in Fig. \ref{Fig1}(b). 
Let's assume the following solution for the elasto-dynamic Eq. \ref{eq_01}:
\begin{equation}
{\bm x}=\frac{1}{2}\left[{\bm a}\left(t\right){\rm e}^{{\rm i}\omega_1\left(t\right)t}+{\bm a}^*\left(t\right){\rm e}^{-{\rm i}\omega_1\left(t\right)t}\right]
\label{eq_05}
\end{equation}
where ${\bm a}=\left[a_1,a_2\right]^T$ are the complex envelopes of oscillators' displacement and $\omega_{1}\left(t\right)=\Omega_1^D\left(t\right)\omega_0$ the temporal evolution of $\Omega_1^D$ during the transformation. A similar relation holds for $\omega_{2}\left(t\right)=\Omega_2^D\left(t\right)\omega_0$, while for ease of visualization, the time dependence of $\omega_1$ and $\omega_2$ in the derivation is implicitly assumed. Differentiating Eq. \ref{eq_05} with respect to time:
\begin{equation}
\begin{split}
\ddot{\bm x}&=\frac{1}{2}\left[{\bm c}\left(t\right){\rm e}^{{\rm i}\omega_1\left(t\right)t}+{\bm c}^*\left(t\right){\rm e}^{-{\rm i}\omega_1\left(t\right)t}\right]\hspace{0.5cm} with:\\[5pt]
{\bm c}\left(t\right)&=\ddot{\bm a}-{\rm i}2\dot{\bm a}\left(\omega_1+\dot{\omega}_1t\right)-{\bm a}\left({\rm i}2\dot{\omega}_1+\left[\omega_1+\dot{\omega}_1t\right]^2\right)
\end{split}
\end{equation}
It is now considered a paraxial approximation of the equation of motion, thus neglecting higher order derivatives, yielding $\dot{\omega}_1t<<\omega_1$, $\dot{\bm a}_{1}<<\omega_{1}{\bm a}_{1}$, and $\ddot{\bm a}_{1}<<\dot{\omega_1}{\bm a}_1+\omega_1\dot{\bm a}_1$, which means that the variation of $\omega_{1}$ and ${\bm a}_1$ belongs to a slower time scale:
\begin{figure*}[t!]
	\centering
	\hspace{-0.4cm}
	\subfigure[]{\includegraphics[width=0.34\textwidth]{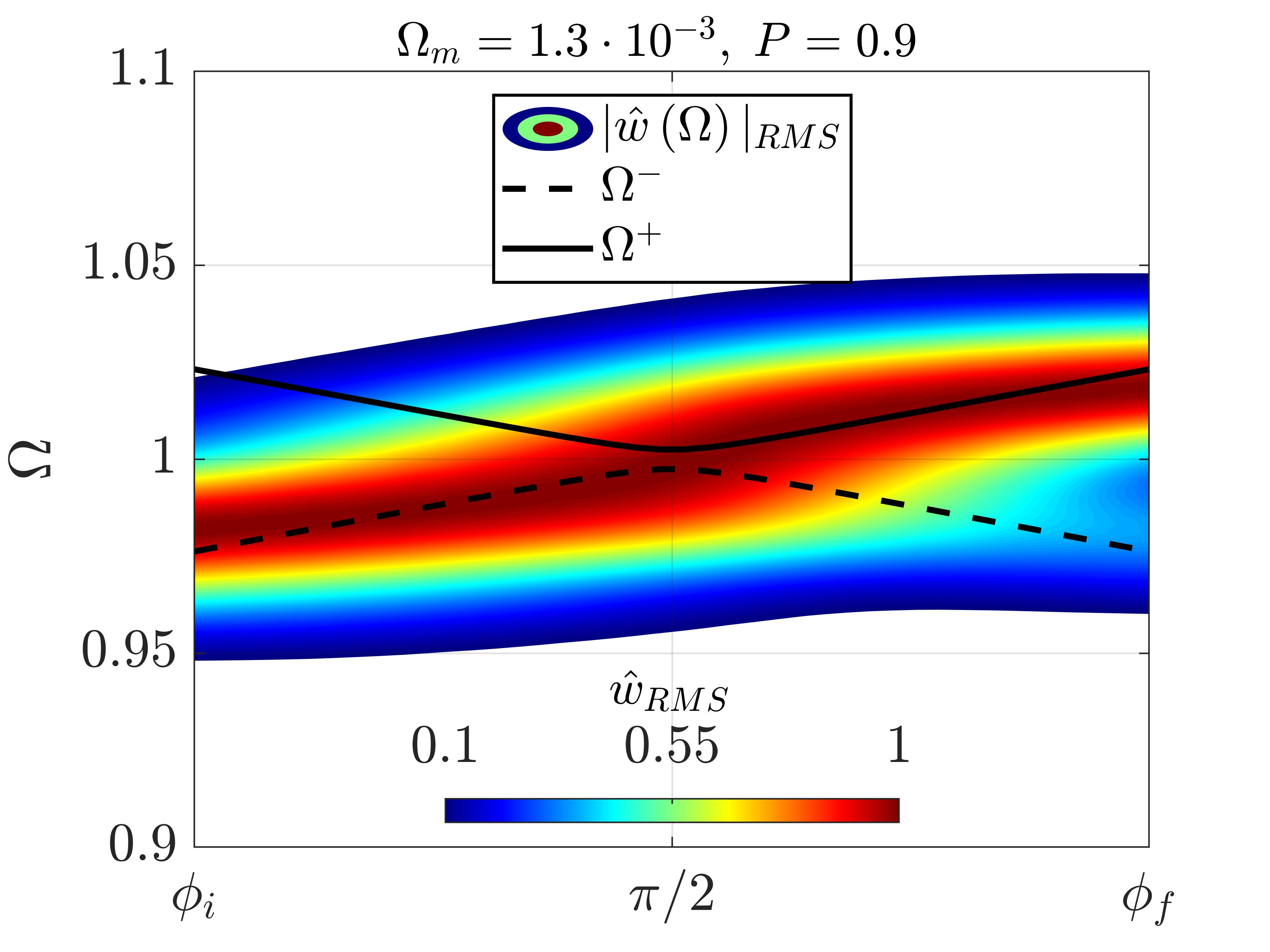}\label{Fig1a}}
	\hspace{-0.4cm}
	\subfigure[]{\includegraphics[width=0.34\textwidth]{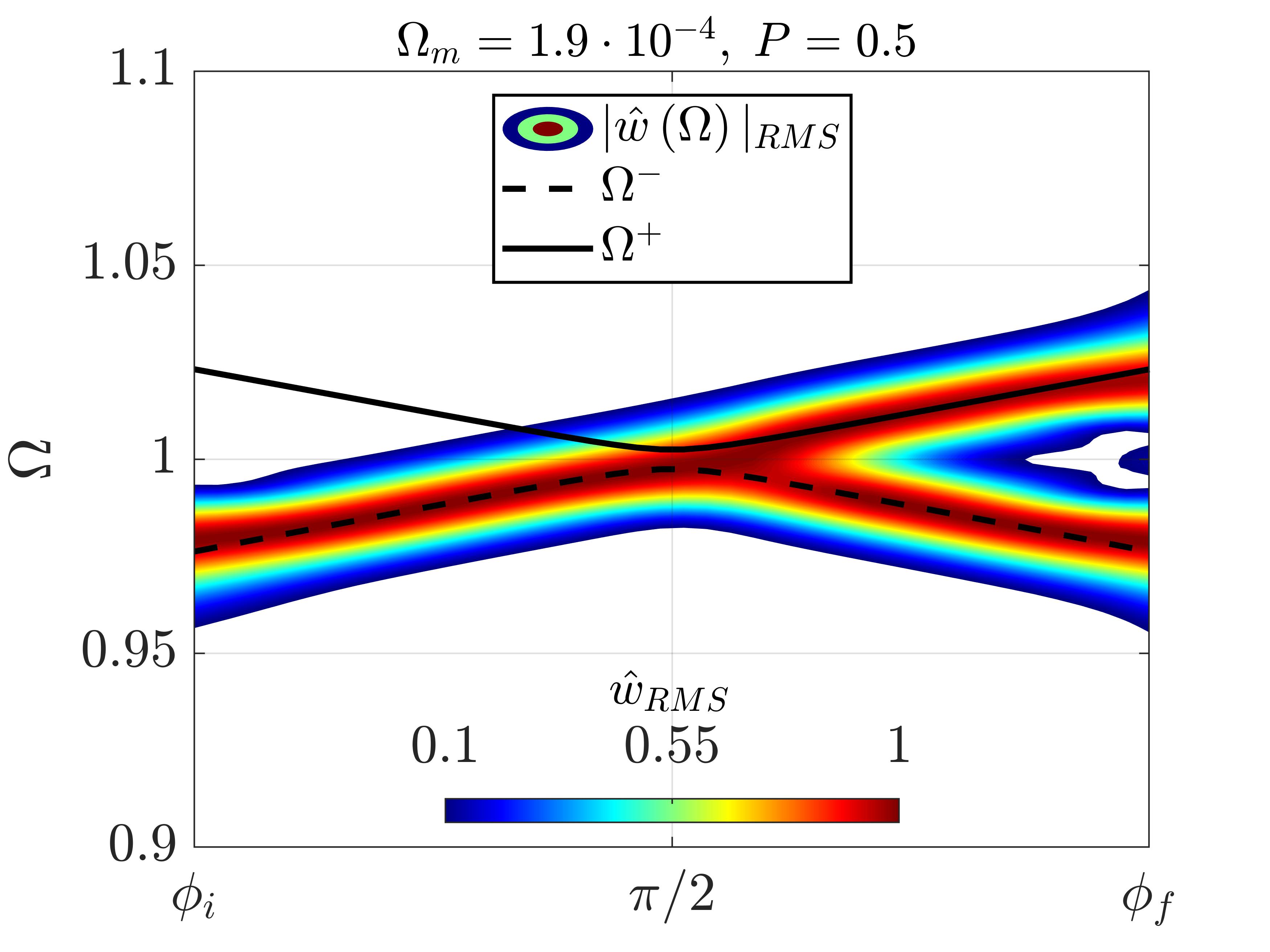}\label{Fig1a}}
	\hspace{-0.4cm}
	\subfigure[]{\includegraphics[width=0.34\textwidth]{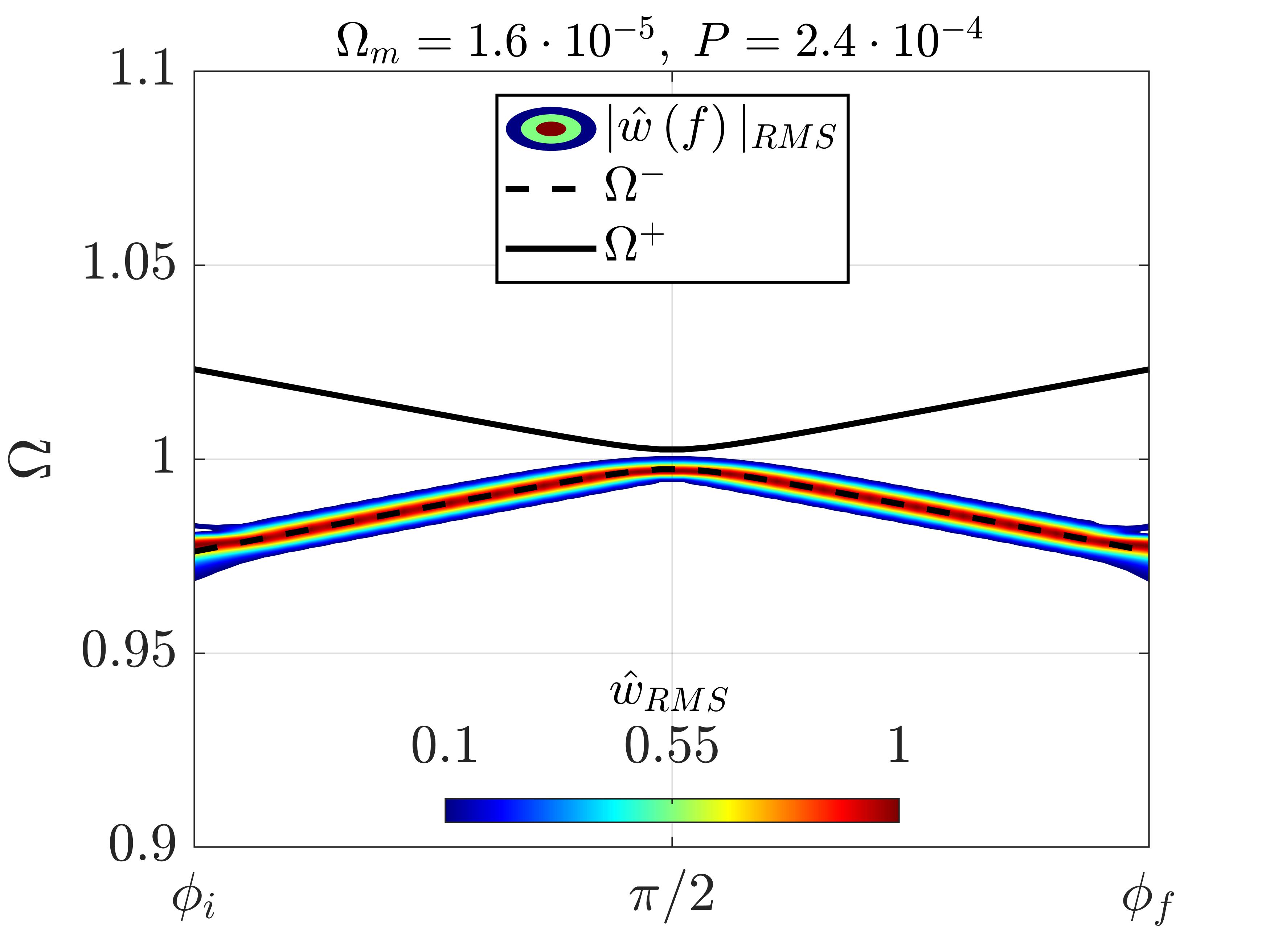}\label{Fig1a}}\\
	\subfigure[]{\includegraphics[width=0.34\textwidth]{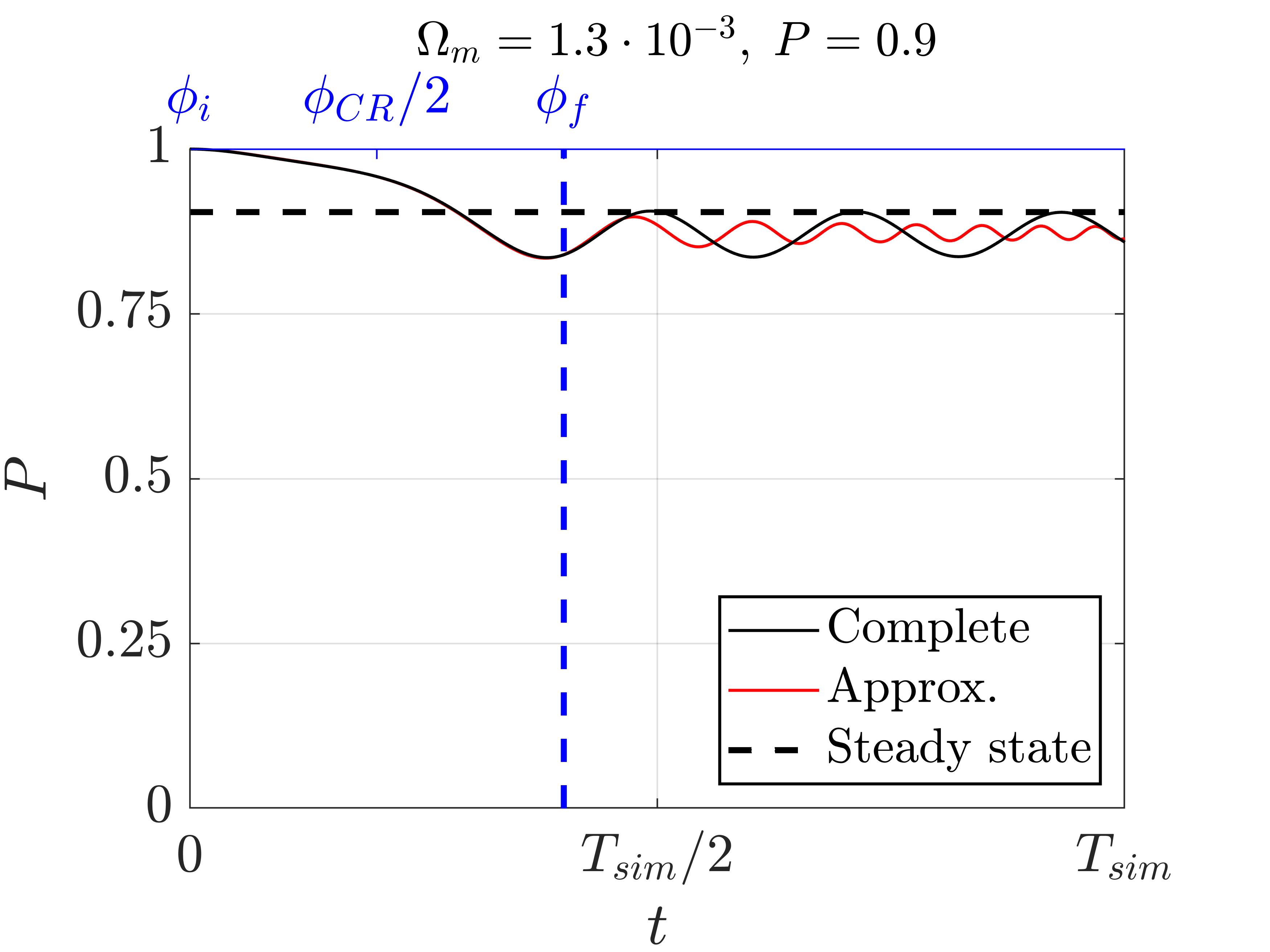}\label{Fig1a}}
	\hspace{-0.4cm}
	\subfigure[]{\includegraphics[width=0.34\textwidth]{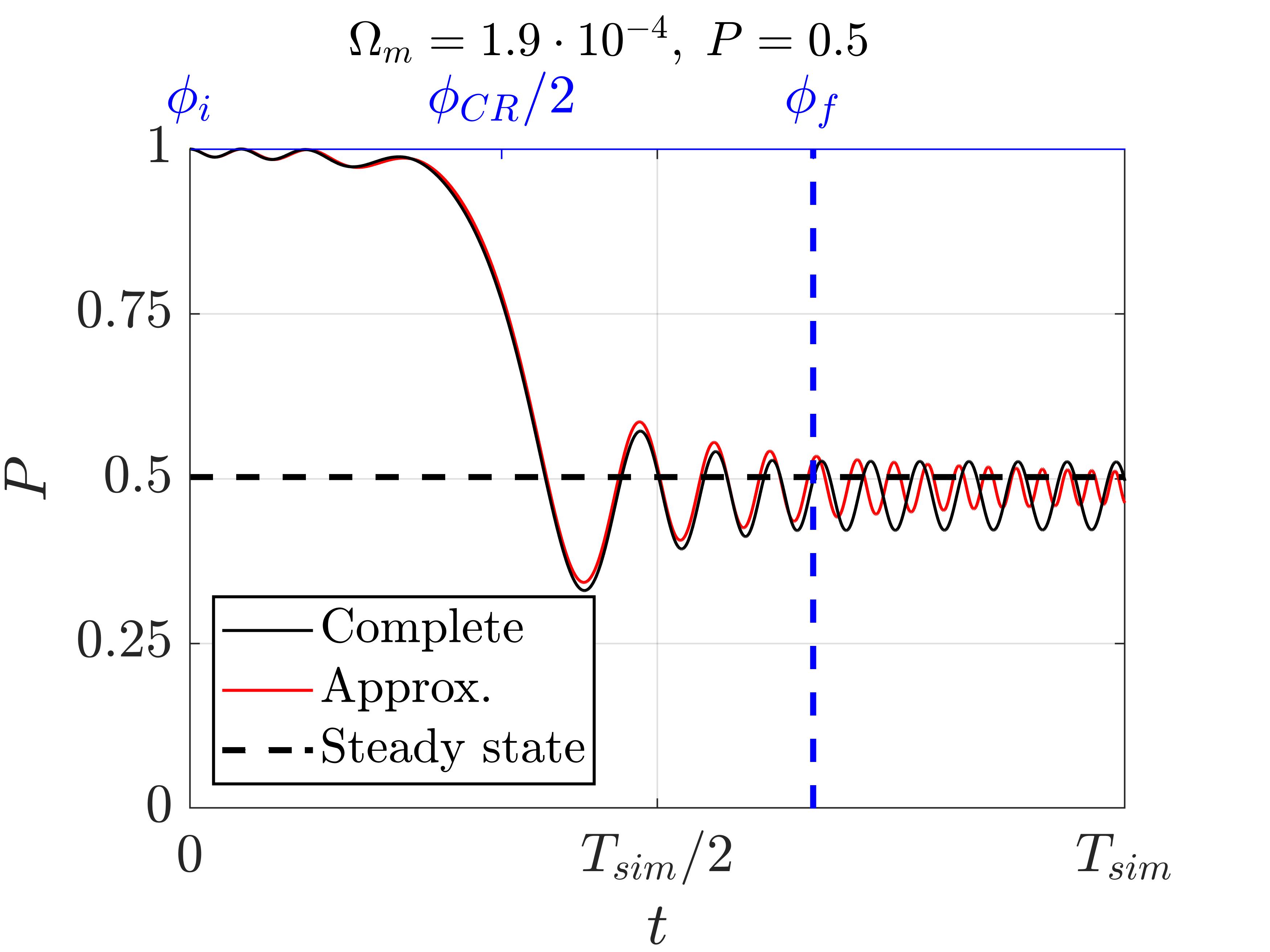}\label{Fig1a}}
	\hspace{-0.4cm}
	\subfigure[]{\includegraphics[width=0.34\textwidth]{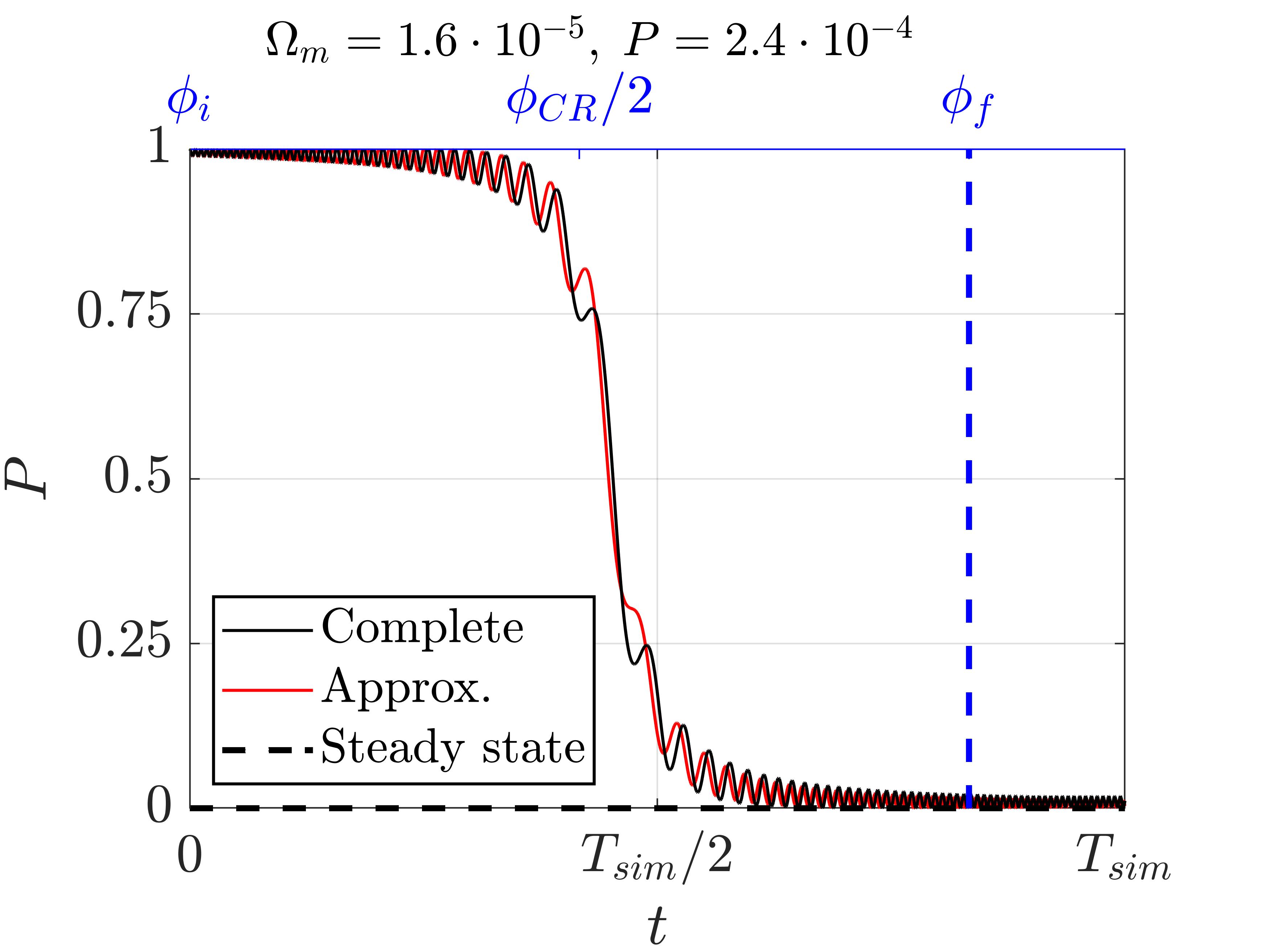}\label{Fig1a}}		
	\caption{(a-c) spectrograms $|\hat{w}\left(\Omega,t\right)|_{RMS}$ of the simulated spring mass system under narrowband spectrum tone burst excitation adopting three different modulation speeds $\Omega_m$. (a) Fast modulation $\Omega_m=1.3\cdot10^{-3}$. (b) Intermediate speed $\Omega_m=1.9\cdot10^{-4}$ and (c) slow modulation $\Omega_m=1.6\cdot10^{-5}$. (d-f) Estimated probability for a state belonging to $\Omega_-^A$ to keep the same polarization and jump to $\Omega_+^A$ in correspondence of $\phi_{CR}=\pi/2$ for distinct modulation speed values. The black curve represents the dynamic probabilities without paraxial approximation of the equation of motion. Red curve: probability time history through paraxial approximation. The asymptotic solution is illustrated with black dashed lines.}
	\label{Fig2}
\end{figure*}	
\begin{equation}
{\bm c}\left(t\right)=-{\rm i}2\dot{\bm a}\omega_1-{\bm a}\omega_1^2
\label{eq_07}
\end{equation}
Plugging Eq. \ref{eq_05}-\ref{eq_07} into Eq. \ref{eq_01} yields the following dynamical system akin to the Schrondinger equation:
\begin{equation}
\begin{split}
&-2{\rm i}\omega_1\dot{a}_1=\Gamma\sqrt{\omega_1\omega_2}a_2\\[5pt]
&-2{\rm i}\omega_1\dot{a}_2-\omega_1^2a_2=\Gamma\sqrt{\omega_1\omega_2}a_1
\end{split}
\label{eq_06}
\end{equation}
which can be rewritten as a second-order ordinary differential equation with time-varying coefficients, by differentiating the first equation and merging it with the second:
\begin{equation}
\ddot{a}_1+\frac{\dot{\omega}_1+{\rm i}\left[\omega_2^2-\omega_1^2\right]/2}{\omega_1}\dot{a}_1+\frac{\Gamma^2}{4}\frac{\omega_2}{\omega_1}a_1=0
\end{equation}
To further simplify the equation it is assumed that $\dot{\omega}_1\dot{a}_1\approx0$, and we consider that the frequency difference between $\omega_{\pm}$ is a linear function of time, which allows for the following approximations:
\begin{equation}
\begin{split}
\frac{\omega_1^2+\omega_2^2}{2\omega_1}&\;\approx\omega_2-\omega_1\approx\beta t;\hspace{0.5cm}\\
\Gamma^2\frac{\omega_2}{\omega_1}&\;\approx\;\Gamma^2\approx\Gamma^2|_{\pi/2}=\left(\frac{k_{12}}{m\omega_{CR}}\right)^2
\end{split}
\end{equation}
where $\omega_{CR}$ is the diabatic frequency evaluated in the crossing point $\phi_{CR}=\pi/2$. $\beta$ is proportional to the modulation velocity:
\begin{equation}
\beta=\frac{d\left(\omega_2-\omega_1\right)}{dt}=\omega_m\gamma
\end{equation}
and the term $\gamma=\partial\left(\omega_2-\omega_1\right)/\partial \phi$ is a constant that approximates linear behavior of the diabatic states in the neighborhood of $\omega_{CR}$. We finally get to:
\begin{equation}
\ddot{a}_1+{\rm i}\beta t\dot{a}_1+\frac{\Gamma_{CR}^2}{4}a_1=0
\label{eq_12}
\end{equation}
where $\Gamma_{CR}$ is $\Gamma$ evaluated in correspondence of the crossing point and represents the frequency separation between states for $\phi=\phi_{CR}$.
The probability function $P\left(t\right)=|a_1\left(t\right)|^2$ for the energy to belong to $\Omega_-^A$ is quantified through Eq. \ref{eq_12}, for a given initial condition $[a_1\left(t_0\right),a_2\left(t_0\right)]$, knowing that $|a_1\left(t\right)|^2+|a_2\left(t\right)|^2=1$.
Also, seeking asymptotic solutions for $t\rightarrow\infty$ yields a constant probability $P\left(t\rightarrow\infty\right)={\rm e}^{-\frac{\pi}{2}\frac{\Gamma_{CR}^2}{\beta}}$, which defines the energy distribution between the states $\Omega_{\pm}$ at the end of the transformation. \\
Finally, assuming that at the initial time instant the energy is entirely located in the bottom state, we enforce initial conditions for $a_1$ to be $a_1(t_0)=1$, which allows for a numerical solution of Eq. \ref{eq_12} in terms of temporal evolution of transition probabilities. \\
We complete the first part of the manuscript numerically solving Eq. \ref{eq_01} with $\phi_i=0.45\pi$, $\alpha=0.3$, $m=1$, $k_0=1$, $\varepsilon=0.005$ and upon comparison between the numerically computed time histories $\bm{x}\left(t\right)$ with respect to the corresponding probabilities. Specifically, the system is excited using a narrowband tone burst excitation for a sufficiently long time period $T$ with a force $F=[1,0]^T\sin(\Omega_-^A\left(t_i\right)\omega_0t)(1-\cos(2\pi/Tt))$ having central frequency $\Omega_-^A\left(t_i\right)$ computed at initial time $t_i$, in order to excite only the state belonging to the bottom branch. After the energy is injected to the target state $\Omega_-^A\left(t_i\right)$, three distinct smooth modulations $\phi=\phi_i+\omega_mt$ are imposed enforcing $\Omega_m=1.3\cdot10^{-3},\;1.9\cdot10^{-4},$ and $1.6\cdot10^{-5}$, corresponding to probabilities of $P=0.9,\;0.5,$ and $2.4\cdot10^{-4}$ respectively, which are evaluated inverting the probability function:
\begin{equation}
\Omega_m=-\frac{\pi}{2\omega_0}\frac{\Gamma_{CR}^2}{\gamma\log\left(P\right)}
\label{eq:omega}
\end{equation} 
Consistently with prior works \cite{nassar2018quantization}, Eq. \ref{eq:omega} illustrates a relationship between the modulation velocity $\Omega_m$, the slope $\gamma$, and frequency separation $\Gamma_{CR}$ of the avoided crossing, for a given probability $P$. That is, the higher the frequency separation, the faster the modulation speed can be for a generic value of $P$. \\
The associated spectrograms are computed employing a fourier transform of the displacement field in reciprocal space $\hat{\bm x}\left(\omega,\kappa_x,t\right)$, by properly windowing the temporal history using a moving Gaussian function \cite{rosa2019edge}. For ease of visualization, the second dimension is eliminated by considering the RMS value along $\kappa_x$. The spectrograms (see Figs. \ref{Fig1}(a-c)) are in good agreement with respect to the steady-state and temporal probabilities displayed in Figs. \ref{Fig2}(d-f) which, in turn, well describe the transitions occurring through the phase modulation. We remark that, for a better visualization of the steady state probability value, the time simulation duration is increased to $T_{sim}$, whereby the final phase modulation time $T_f=(\phi_f-\phi_i)/\omega_m$ is highlighted with a vertical blue line.
Specifically, Fig. \ref{Fig2}(a,d) display a fast transition with frequency shift and without eigenvector transformation, which is consistently described by the probability $P=0.9$ for a state to keep the same polarization and therefore to jump from $\Omega_-^A$ to $\Omega_+^A$.  
Fig. \ref{Fig2}(b,e) instead describe frequency splitting, in which half ($P=0.5$) of the energy remains to $\Omega_-^A$ and half jumps to $\Omega_+^A$. Finally, Fig. \ref{Fig2}(c,f) illustrate an almost adiabatic transition with eigenvector transformation, in which the starting state remains located at the bottom branch $\Omega_-^A$, which yields the almost zero probability for the state to keep the initial polarization.\\
It is worth mentioning that the steady state probability $P\left(t\rightarrow\infty\right)$ and the corresponding time history obtained from numerical integration of Eq. \ref{eq_12} exhibit a small difference in the steady-state behavior, especially when the numerical integration duration is short.  This mismatch results from the different time domains considered for computing the aforementioned solutions which, in one case is $[0,\infty]$ and in the second case is $\left[0,T_f\right]$.
\section{Edge-to-edge pumping in a quasiperiodic beam}
\begin{figure*}[t!]
	\centering
	\hspace{-0.4cm}
	\subfigure[]{\includegraphics[width=0.54\textwidth]{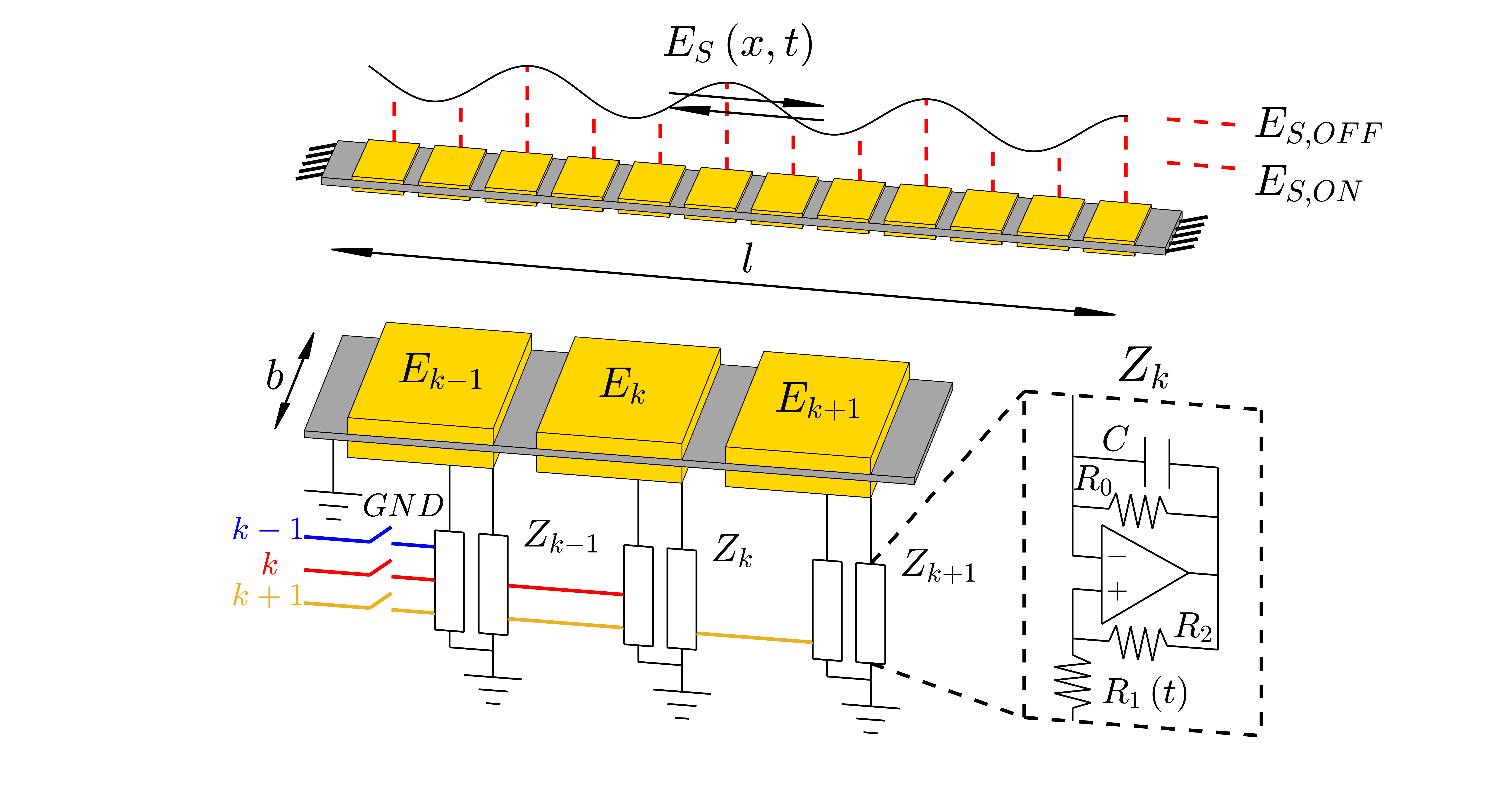}\label{Fig1a}}
	\hspace{-0.4cm}
	\subfigure[]{\includegraphics[width=0.38\textwidth]{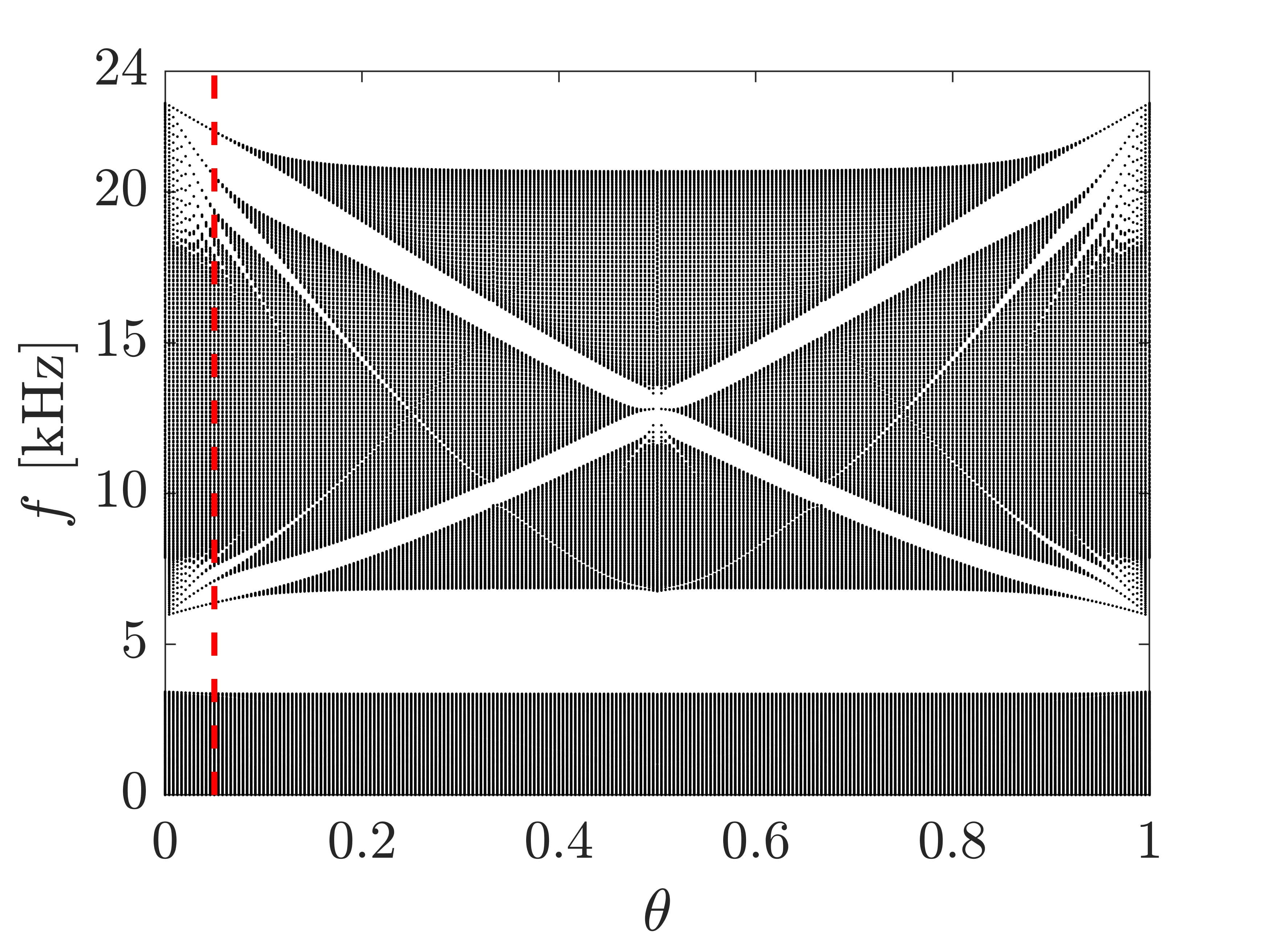}\label{Fig1a}}\\
	\subfigure[]{\includegraphics[width=0.32\textwidth]{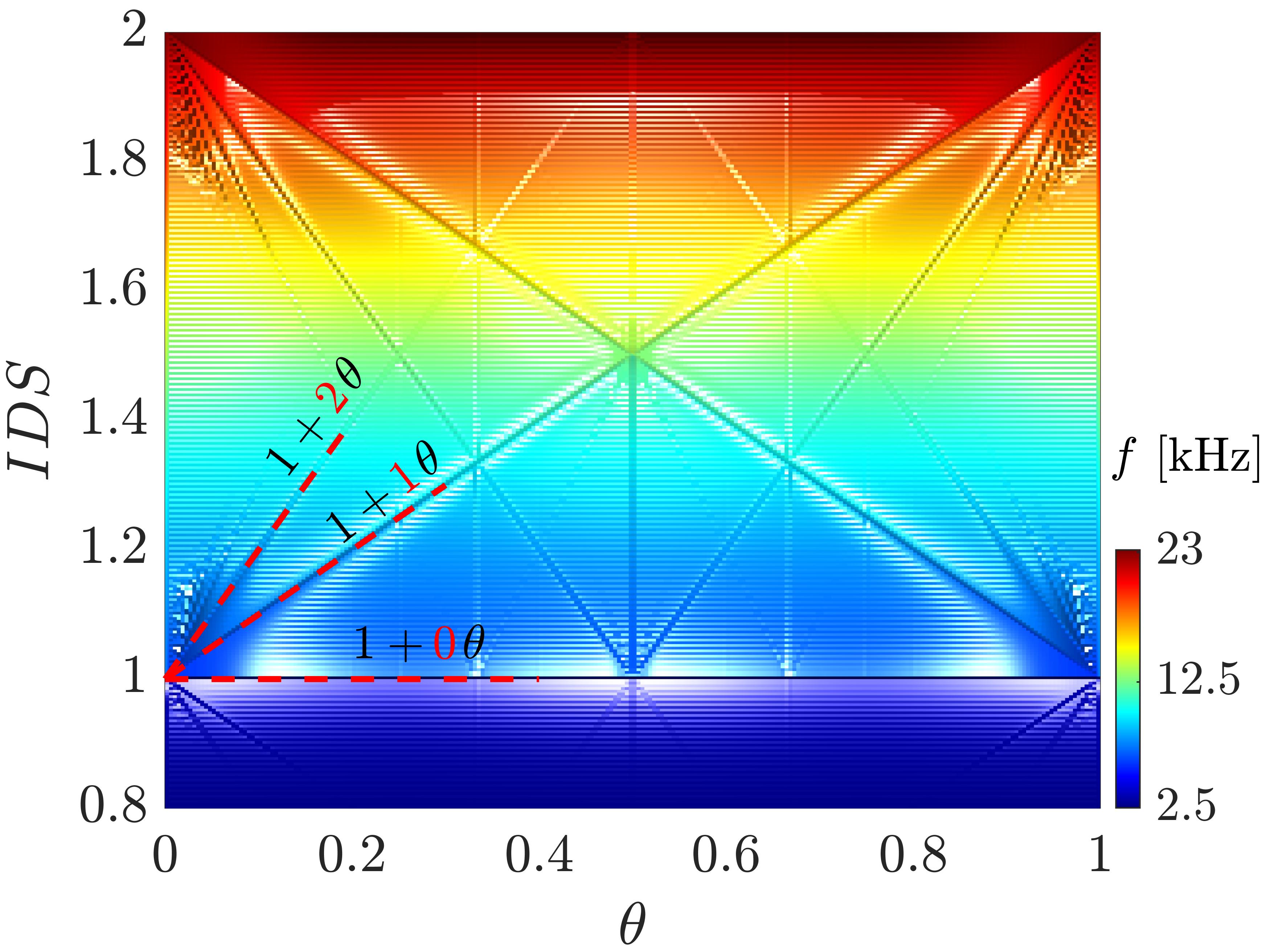}\label{Fig1a}}
	\subfigure[]{\includegraphics[width=0.33\textwidth]{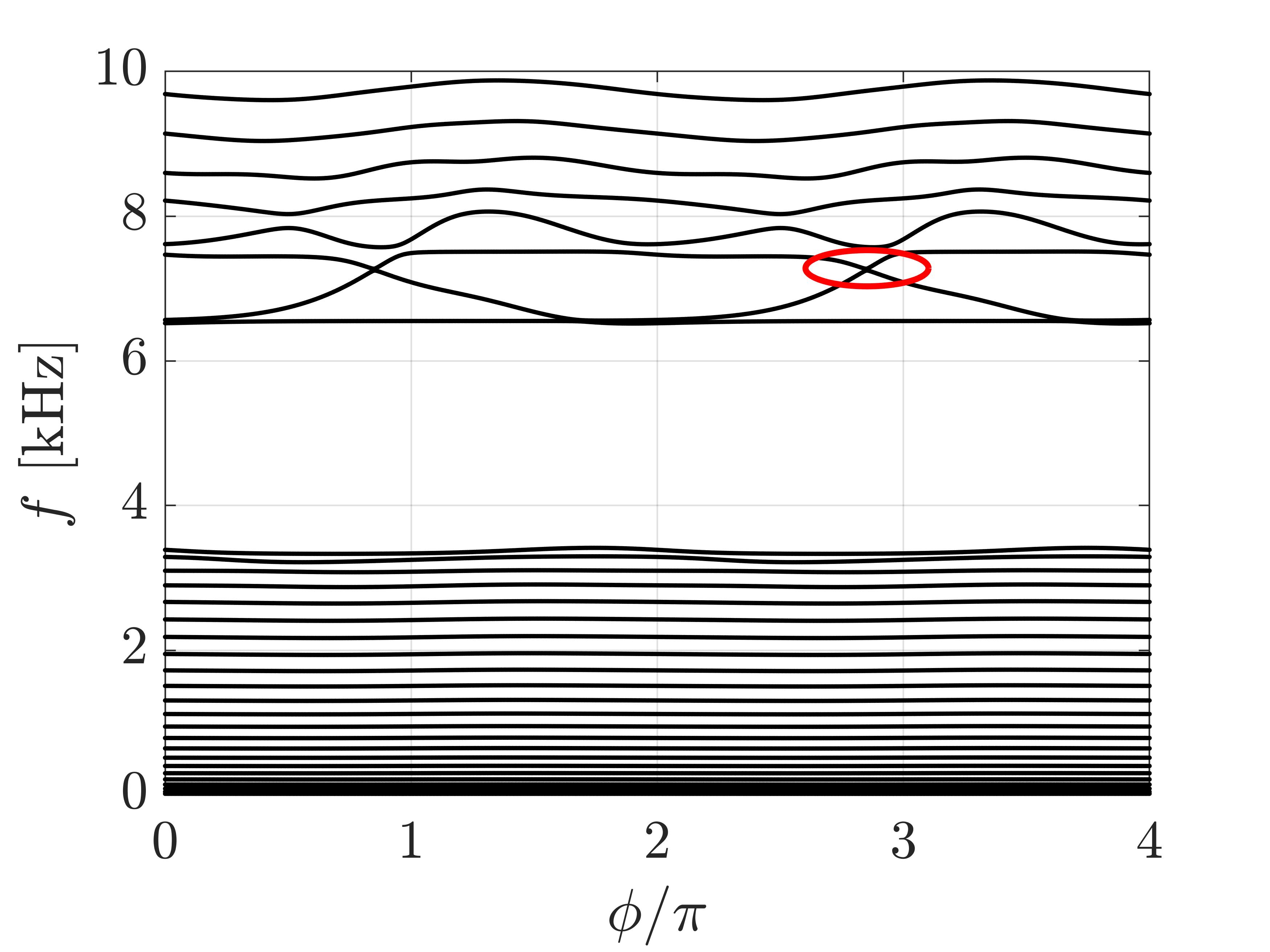}\label{Fig1a}}
	\subfigure[]{\includegraphics[width=0.33\textwidth]{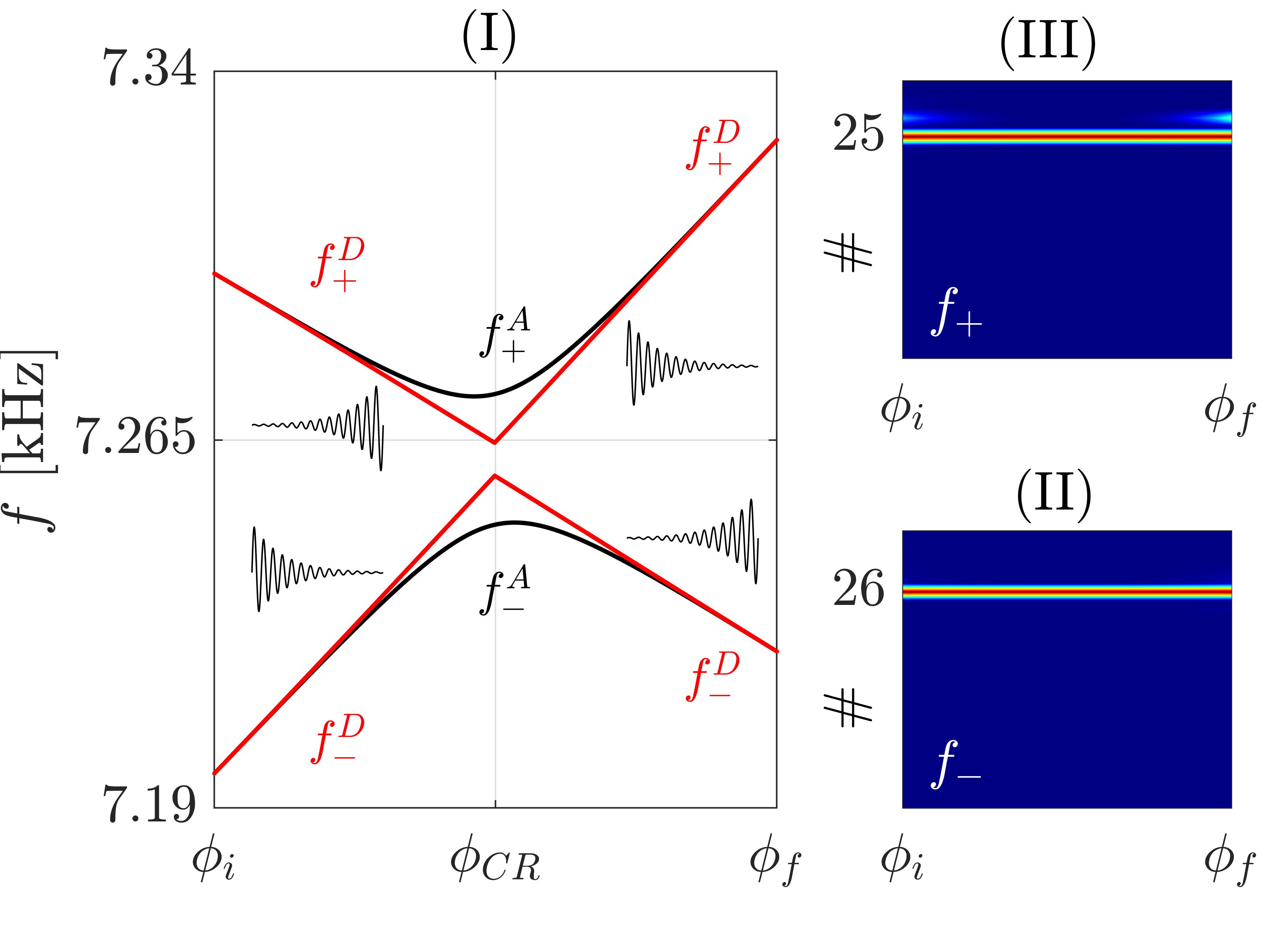}\label{Fig1a}}
	\caption{Schematic of the beam under clamped-clamped boundary conditions and quasiperiodic stiffness modulation. Quasiperiodicity is achieved by means of non-repeating control signals $k-1,\;k,\;k+1,\ldots$ able to locally alter the electrical parameters of the NC shunt. In the schematic, a temporal modulation of $R_1$ is assumed. (b) Hofstadter butterfly associated to a commensurate realization of a stiffness modulated beam upon varying the projection parameter $\theta$. The vertical dashed line corresponds to the configuration adopted to study edge-to-edge transitions. (c) Integrated Density of States (IDS). The slope of the highlighted lines yield the labels of the gap, which are written in red. (d) Spectrum of the system for $\theta = 0.075$ and upon varying the modulation phase $\phi$. A pair of states is spanning the gap and generate the \textit{avoided crossing}.  (e) $I$ Zoomed view of the avoided crossing. Adiabatic states in black and approximated diabatic states in red. In the figure, the schematic of the expected polarization for each branch is illustrated. (e) ${\rm II-III}$ Estimated \textit{Modal Dependence Factor} (MDF) between modes $\#25,\;\#26$ and full spectrum.}
	\label{Fig3}
\end{figure*}
Consider now a real and application-oriented case-study, in which a plain aluminum beam is equipped with periodically placed smart piezoelectric patches, for a total of $N=24$ pairs bonded on the top and bottom surfaces. The coupling between electrical and mechanical domains enables stiffness modulation when subjected to certain electrical boundary conditions which, in the case at hand, are \textit{negative capacitance} (NC) shunts. In addition, the circuit's components are temporally modulated in time providing effective Young's modulus variation according to a predetermined modulation law. Such configuration has been successfully employed in prior studies concerning space-time modulations \cite{marconi2019experimental} and shown in Fig. \ref{Fig3}(a). The corresponding physical and geometrical properties are reported in Appendix A. Let's assume that consecutive sub-elements are stiffness modulated in the following fashion:
\begin{equation}
E_k=E_{s,0}\left[1+\alpha\cos{\left(2\pi\theta k + \phi\left(t\right)\right)}\right]
\label{eq_13}
\end{equation}
being $E_{s,0}$ the mean effective Young's modulus of the sandwich structure, $\alpha$ a dimensionless modulation amplitude  and $\phi\left(t\right)=\phi_i+\omega_mt$ is a phase parameter which is a smooth function of time. $k$ denotes the $k^{th}$ piezo pair along the beam's main dimension.
\begin{figure*}[t!]
	\centering
	\hspace{-0.4cm}
	\subfigure[]{\includegraphics[width=0.34\textwidth]{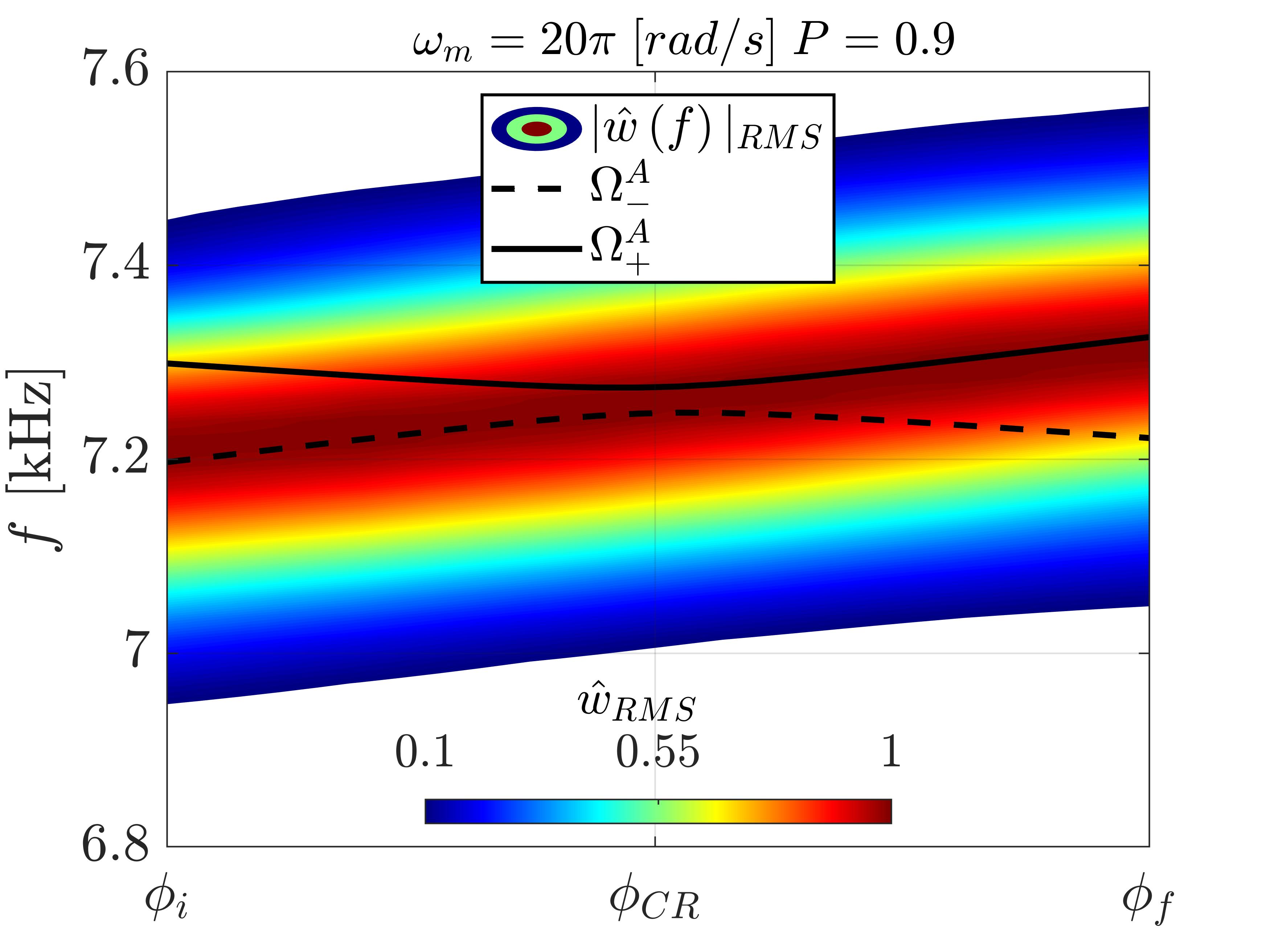}\label{Fig1a}}
	\hspace{-0.4cm}
	\subfigure[]{\includegraphics[width=0.34\textwidth]{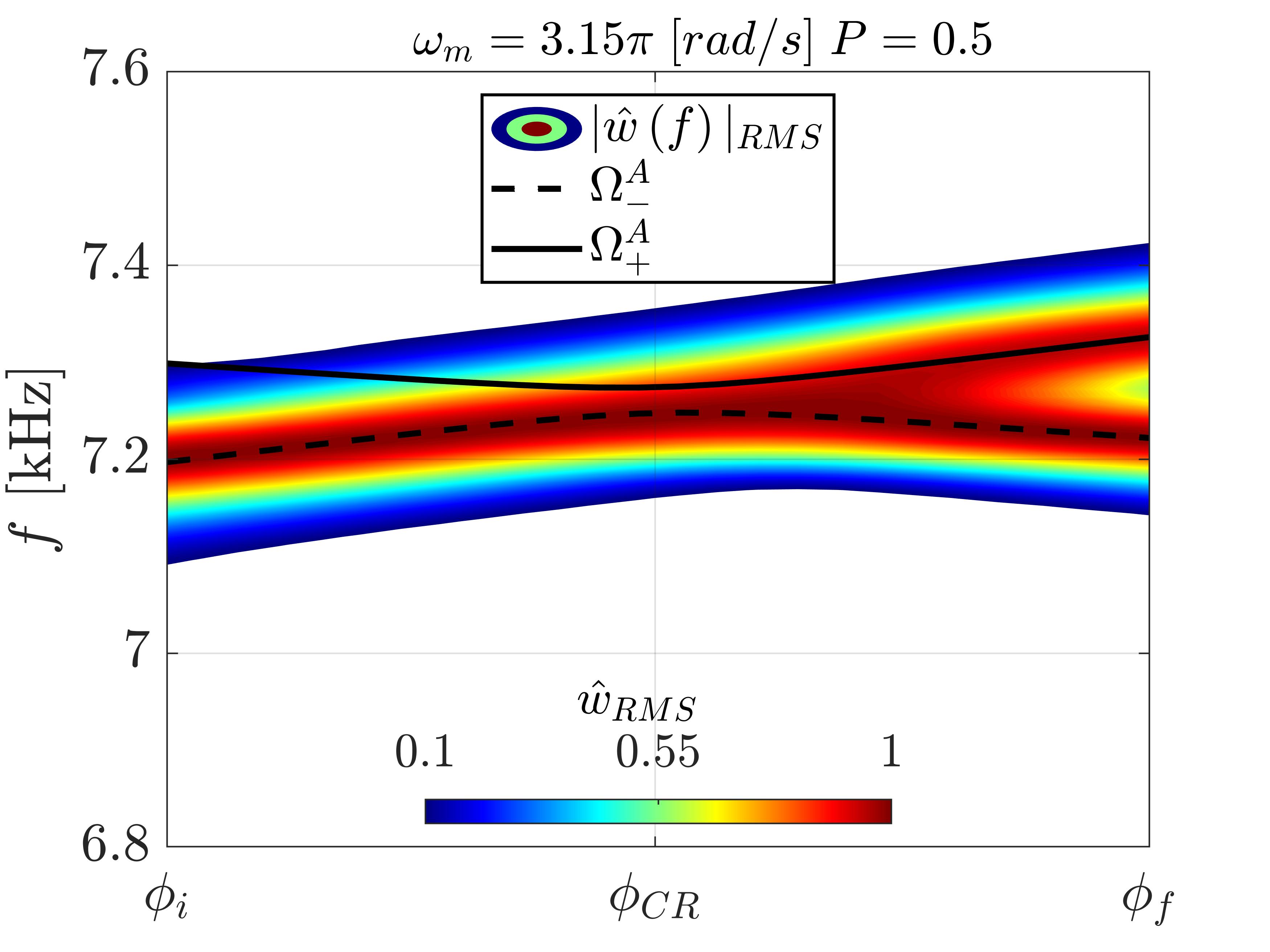}\label{Fig1a}}
	\hspace{-0.4cm}
	\subfigure[]{\includegraphics[width=0.34\textwidth]{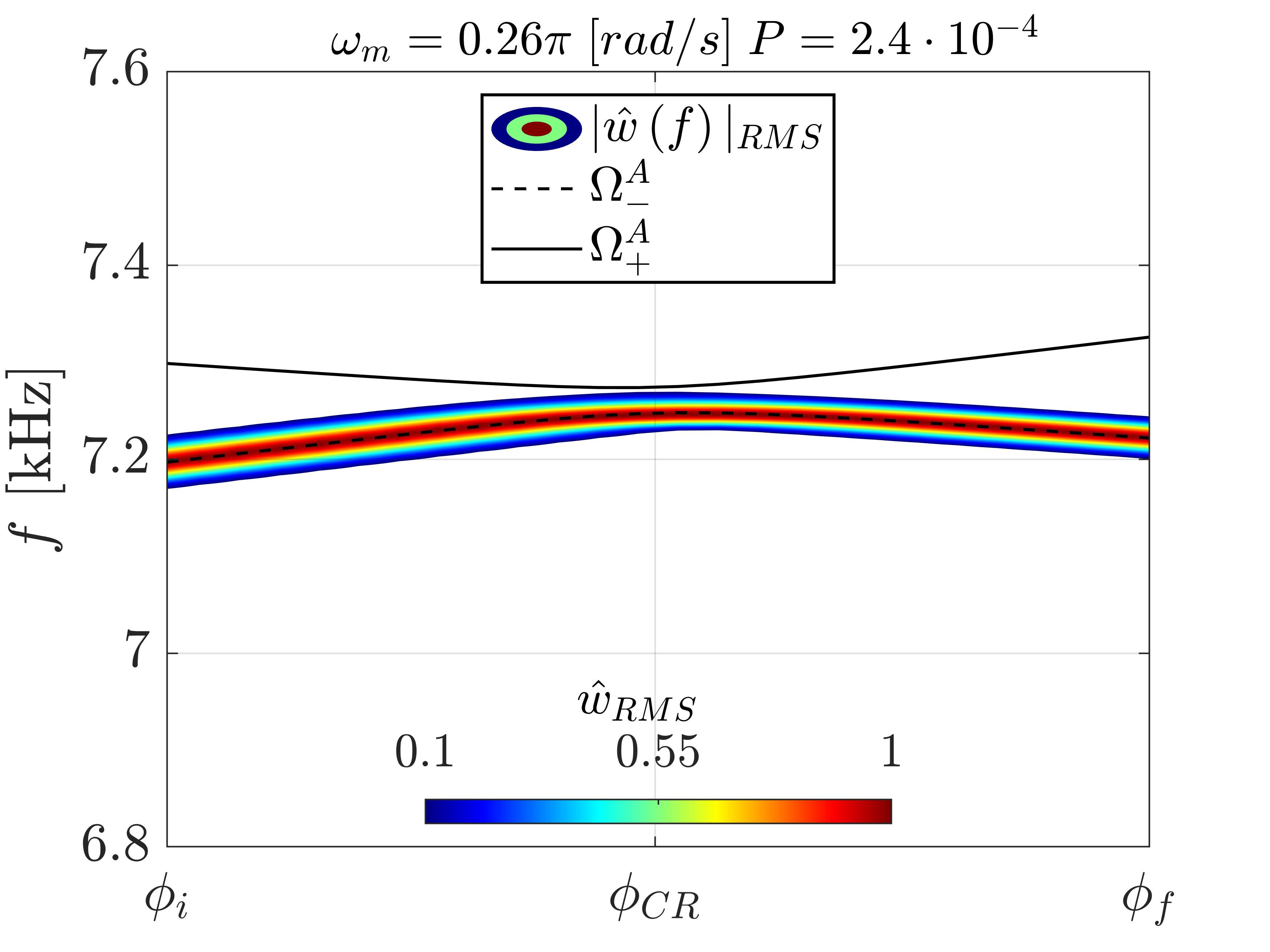}\label{Fig1a}}
	\hspace{-0.4cm}
	\subfigure[]{\includegraphics[width=0.34\textwidth]{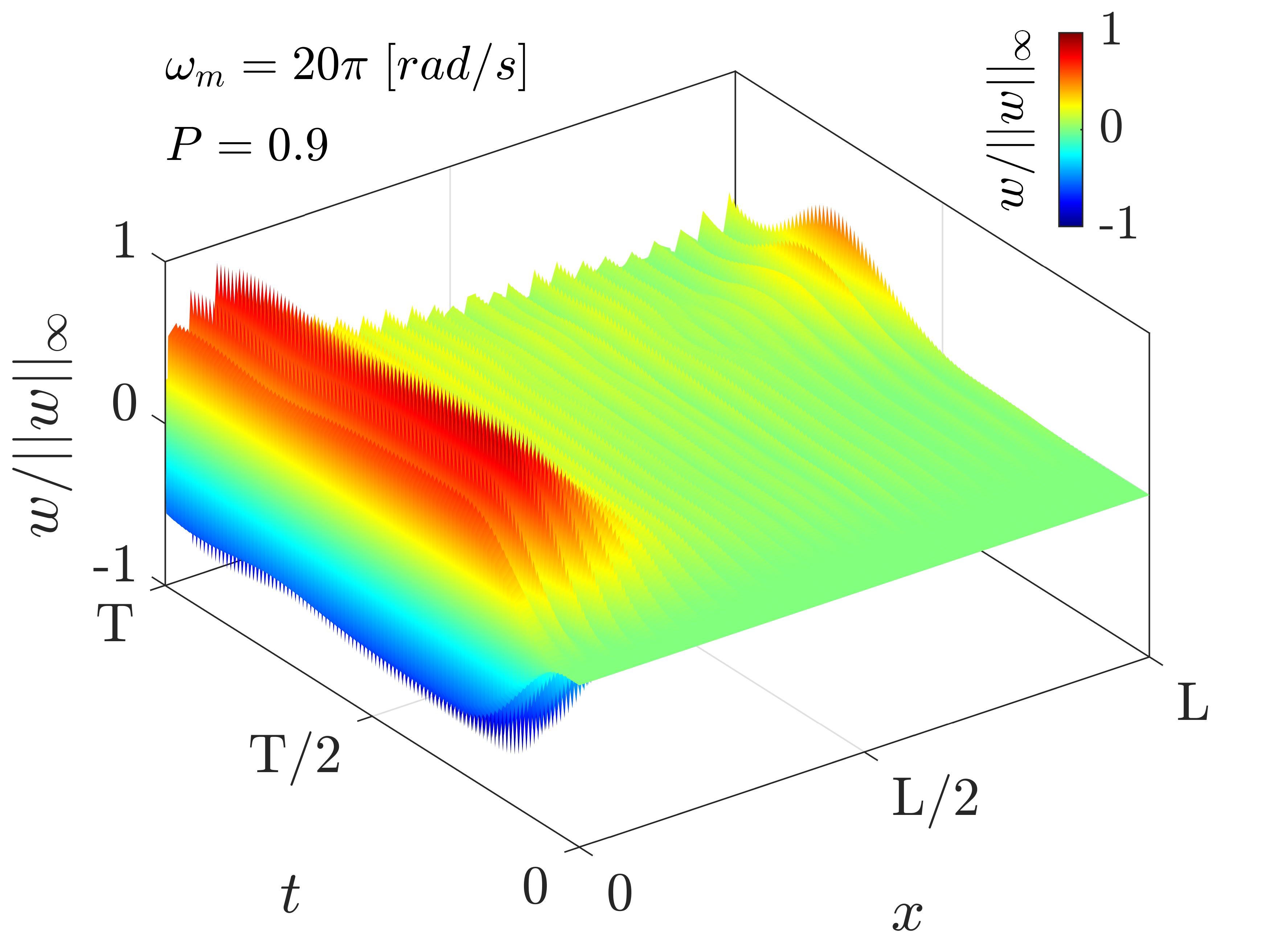}\label{Fig1a}}
	\hspace{-0.4cm}
	\subfigure[]{\includegraphics[width=0.34\textwidth]{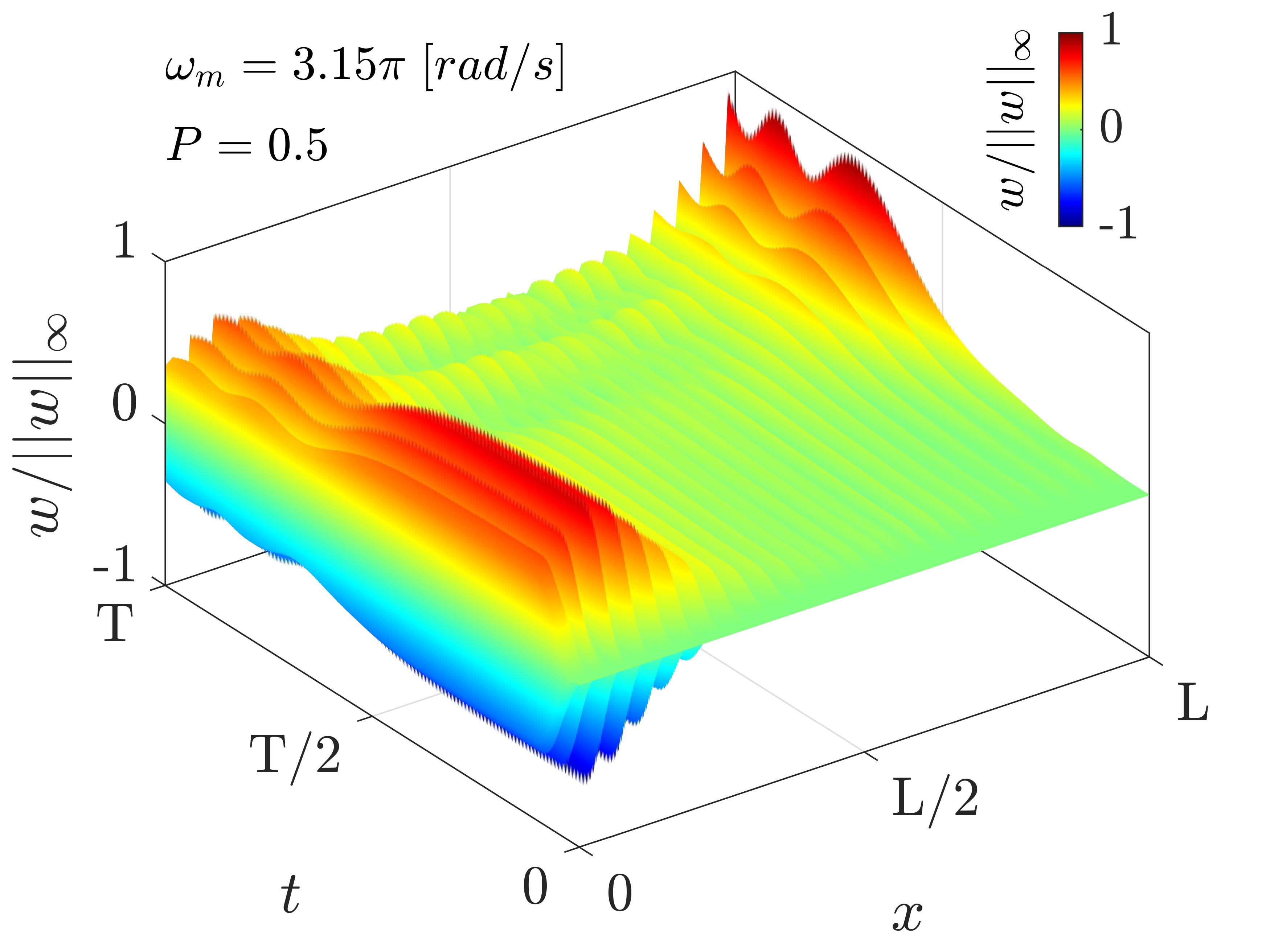}\label{Fig1a}}
	\hspace{-0.4cm}
	\subfigure[]{\includegraphics[width=0.34\textwidth]{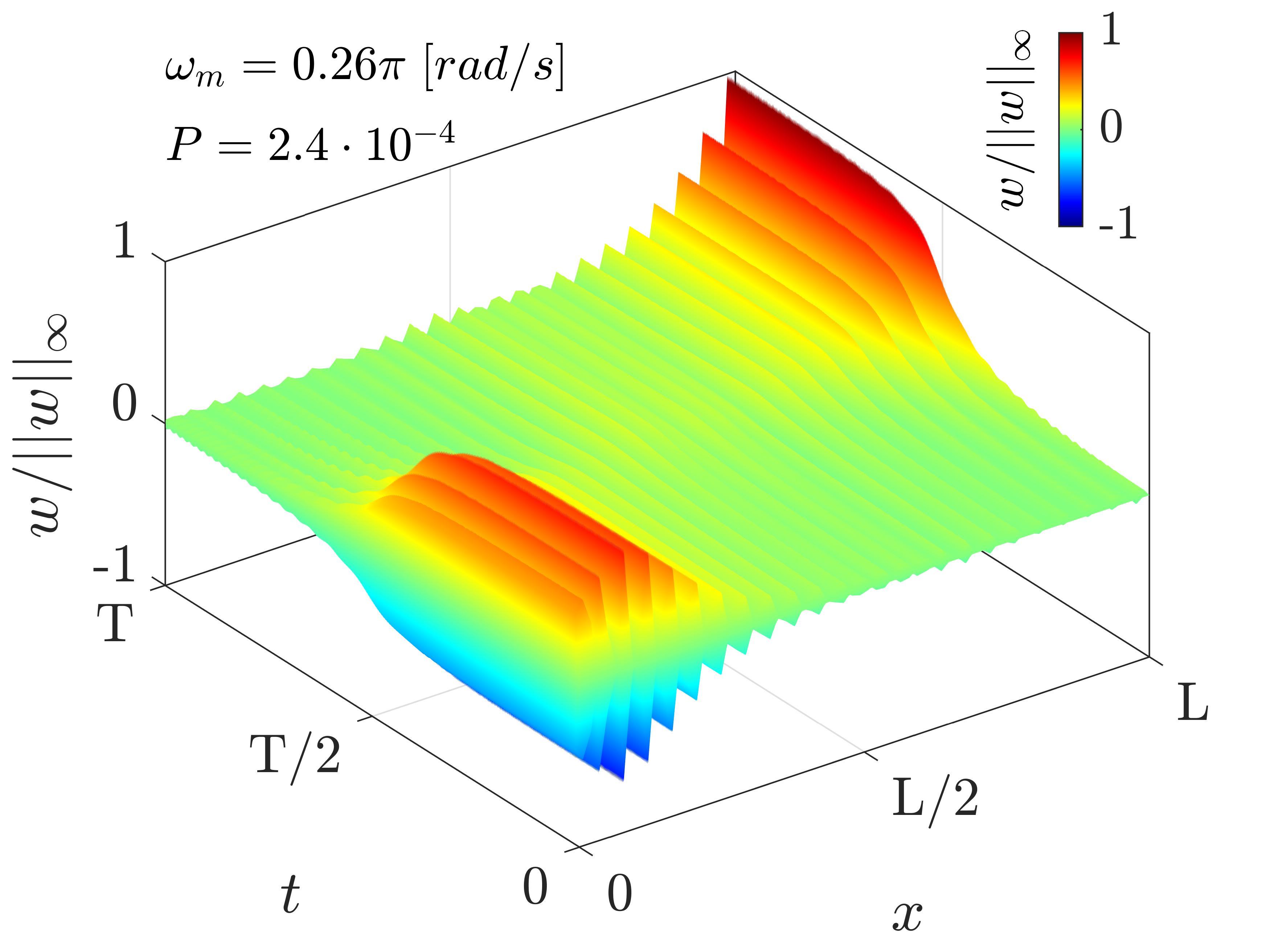}\label{Fig1a}}
	\caption{(a-c) Spectrograms $|\hat{w}\left(f,t\right)|_{RMS}$ resulting from the numerical time history obtained through narrowband burst excitation of the quasiperiodic beam. The shape of the force is tailored to favor the excitation of the left-localized mode. The modulation phase is varied with three different velocity levels $\omega_m$. (a) Fast modulation $\omega_m=20\pi$. (b) Intermediate speed $\omega_m=3.15\pi$. (c) Slow modulation $\omega_m=0.26\pi$. (d-f) Corresponding displacement field in space and time. (d) without edge-to-edge transition. (e) with $50\%$ energy left localized and $50\%$ right localized. (f) Complete edge-to-edge pumping.}
	\label{Fig4}
\end{figure*}
Interestingly, such modulation embodies nontrivial topological properties, which reflects into a fractal spectrum associated to $\theta$ variations for a commensurate realization of the beam, as shown through the \textit{Hofstadter butterfly} in Fig. \ref{Fig3}(b). 
The nontrivial nature of the gaps is quantified through a graphical interpretation of the \textit{Integrated Density of States} (IDS), which is illustrated in Fig. \ref{Fig3}(c), whereby the labels for the gap $C_g=m$ are equivalent to the slope of the red dashed lines, since $IDS=n+m\theta$ for this family of modulations \cite{xia2020topological}.\\
We now focus our attention above the first trivial gap ($C_g=0$), and we employ a quasiperiodic configuration of the system, whose projection parameter $\theta=0.075$ corresponds to the red dashed line in Fig. \ref{Fig3}(b). 
The associated spectrum upon varying the modulation phase $\phi$ is illustrated in Fig \ref{Fig3}(d) and exhibits a first nontrivial gap ($C_g=1$) at approximately $6\;{\rm kHz}$. Interestingly, a pair of topological edge states is observed when cyclic variation of $\phi$ are considered, whose dependence with $\phi$ manifests as \textit{avoided crossing}.
The topological characteristics and localization properties of similar quasiperiodic configurations have been extensively discussed in \cite{rosa2019edge}. Here, instead, we investigate on the avoided crossing dynamics, which is observable within $\left[\phi_i,\phi_f\right]=\left[2.82\pi,2.88\pi\right]$. A zoomed view in the neighborhood of $\phi_{CR}$ is illustrated in Fig \ref{Fig3}(e), corresponding to mode polarizations which are left and right localized for $f_-^A\left(\phi_i\right),\;f_+^A\left(\phi_f\right)$ and $f_+^A\left(\phi_i\right),\;f_-^A\left(\phi_f\right)$ respectively, providing opportunities for edge-to-edge transitions, similarly to section \ref{section2}. \\
In contrast to simple spring-mass systems, the modal coupling and diabatic frequencies are unknown and, for an estimation of the latter, we exploit the numerical procedure previously discussed. To this end, the adiabatic basis ${\bm \Psi}^A_i$ and ${\bm \Psi}^A_f$ are computed through a finite element approximation of the system, yielding the following eigenvalue problem:
\begin{equation}
\left(-\omega^2{\bm M}+{\bm K}\right){\bm w}=0
\end{equation} 
Such basis are then exploited to perform a change of coordinates ${\bm w}={\bm \Psi}^A_{i,f}{\bm q}$ and therefore used for an evaluation of the generalized mass ${\bm M}_q=({\bm \Psi}^A_{i,f})^T{\bm M}{\bm\Psi}^A_{i,f}$ and stiffness ${\bm K}_q=({\bm\Psi}^A_{i,f})^T{\bm K}{\bm\Psi}^A_{i,f}$ matrices. In analogy with section \ref{section2}, the coefficients $K_{q}^{25,26}$ and $K_{q}^{26,25}$ are responsible for the coupling between otherwise degenerate states, whereby setting $K_q^{25,26}=K_q^{26,25}=0$ breaks the modal interactions and converts adiabatic frequencies $f_\pm^A$ to the corresponding diabatic approximations $f_\pm^A\rightarrow f_\pm^D$.
A comparison between $f_\pm^A$ (black curves) and $f_\pm^D$ (red lines) is illustrated in Fig \ref{Fig3}(e), whereby on one hand, the adiabatic curves are converted into diabatic states nullifying the modal coupling. On the other hand, the small gap between estimated states for $\phi=\phi_{CR}$ it is justified by the approximation ${\bm\Psi}^D_{i,f}\approx{\bm\Psi}^A_{i,f}$.
Moreover, an estimation of the coupling between modes $25,26$ and the full spectrum is shown in Fig. \ref{Fig3}(e) ${\rm II-III}$ in terms of \textit{Modal Dependence Factor} (MDF), whose computation is addressed following a procedure detailed in Appendix B. As expected, the state belonging to $f_-^A$ (i.e. mode $\#\;25$) couples only with $f_+^A$ (mode $\#\;26$) through the coupling coefficient $K_q^{25,26}$. While mode $\#\;26$ mainly couples with $\#\;25$ except for from small interactions with $\#\;27$ in correspondence of $\phi_i$ and $\phi_f$.\\
Now, based on the diabatic frequency estimation, the required modulation speed $\omega_m$ is evaluated as a function of the target steady-state transition probabilities $P$, the frequency separation $\Gamma_{CR}$, and the crossing slope $\gamma$, in agreement to Eq. \ref{eq:omega}.
The estimated probabilities and modulation speeds are then validated numerically by solving the adiabatic and non-adiabatic transition problems. Specifically, a narrowband input spectrum, which is able to favor only the excitation of the left localized state, is considered. To this end, the shape of the input force is set coincident to the mode polarization ${\bm w}^A$ of the $25^{th}$ state computed for $\phi=\phi_i$. Later, the phase parameter is varied using three distinct values for $\omega_m=20\pi,\;3.15\pi,0.26\pi\;{\rm [rad/s]}$. The resulting displacement field is used to recover the energy content in reciprocal space $\hat{w}\left(f,\kappa_x,t\right)$, which is then reduced to $\hat{w}\left(f,t\right)$, by taking the RMS value along $\kappa_x$. In case of fast modulation ($\omega_m=20\pi$), the probability for the initial state to keep the same polarization (i.e. to jump from $\Omega_-^A$ to $\Omega_+^A$ for $\phi_{CR}$) is $0.9$, which is confirmed by the associated spectrogram in Fig. \ref{Fig4}(a). Consequently, the temporal evolution of the beam's displacement (see Fig. \ref{Fig4}(d)) illustrates that the topological state remains left localized (except for some energy that leaks to the right), which reflects the mode polarization ${\bm x^A}$ associated to the branch the solution belongs to. 
When the intermediate speed of $\omega_m=3.15\pi$ is applied to the system, corresponding to $P=0.5$, the energy content splits between two states which are left and right localized respectively, as shown in Fig. \ref{Fig4}(b,e). Finally, a slow modulation, characterized by $\omega_m=0.26\pi$ and $P=2.4\cdot10^{-4}$, results in a complete topological transitions from the left to right boundaries, therefore achieving a topological pump. The corresponding specrogram demonstrates that the topological transition occurs with a frequency shift, so that the initial state keeps belonging to the bottom branch with negligible scattering of energy to the neighboring modes.
\section{Conclusions}
In this manuscript it is demonstrated that the coupling between distinct topological states populating the same gap leads to the formation of avoided crossings characterized by edge-to-edge transitions.
The avoided crossing dynamics is investigated in the context of quasiperiodic stiffness modulated beams and, specifically, we have shown a systematic procedure to break the modal coupling upon approximation of the diabatic frequencies and corresponding basis, which is used to estimate the required speed of modulation for a given edge-to-edge transition probability.
The results presented in the paper can be of technological relevance for applications involving elastic energy splitting and demultiplexing, frequency conversion and waveguiding.
\section*{Acknowledgments}
The authors wish to thank Dr. Massimo Ruzzene and M. Rosa for useful discussion in the preliminary stage of the present work.\\
The Italian Ministry of Education, University and Research is acknowledged for the support provided through the Project "Department of Excellence LIS4.0 - Lightweight and Smart Structures for Industry 4.0”.

\appendix
\newcommand{\beginappendix}{%
	\setcounter{table}{0}
	\renewcommand{\thetable}{A\arabic{table}}%
	\setcounter{figure}{0}
	\renewcommand{\thefigure}{A\arabic{figure}}%
}
\beginappendix
\section{Data of the quasiperiodic beam}
{In this manuscript the analysis are performed considering the electroelastic beam illustrated in Fig. \ref{Fig3}(a), which is made of an aluminum substrate with cross section \textit{b} x \textit{H} = 20 mm x 1 mm and total length \textit{l} = $576$ mm. 
An array of piezoelectric patches, separated by a 2 mm distance, consists of 24 piezo-pairs bonded on opposite surfaces with material density \textit{$\rho_p$} = $7.9\;{\rm kg/dm^3}$, short circuit Young's modulus $E_p = 62\; {\rm GPa}$, and size $b$ $\times$ $h_p$ $\times$ $l_p$ = 20 $\times$ 1 $\times$ $22$ mm. The boundary conditions are clamps applied to both beam's ends. \\
Each patch is connected to a shunt circuit emulating a series negative capacitance (NC), for a total of 48 shunts, which provide an effective stiffness decrease to the beam sandwich when the circuit is active \cite{marconi2019experimental}.\\
\begin{figure}[t]
	\centering
	\includegraphics[scale=0.52]{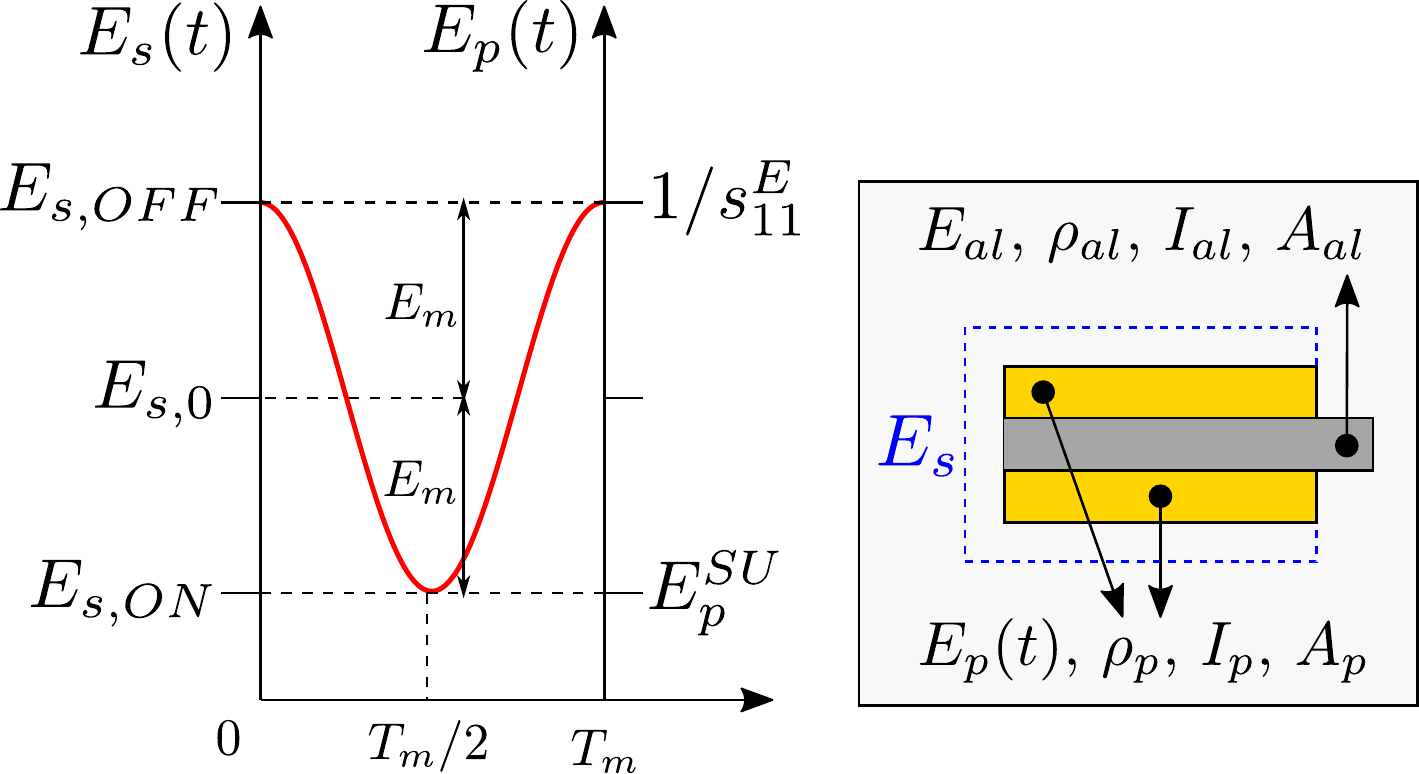}
	\caption{Schematic and adopted notation of stiffness for the patch ($E_p$), the sandwich ($E_s$) and of relevant modulation parameters.}
	\label{Fig5}    
\end{figure}
In the case at hand, the modulation law reflects the electrical boundary conditions applied to the piezoelectric patches in agreement with the circuit schematic in Fig. \ref{Fig3}(a) which, in turn, locally alters the effective Young's modulus of the material in the following fashion:
\begin{equation}
E_p^{SU}=E_p\frac{C_N-C_p^T}{C_N-C_p^S}
\end{equation} 
being $C_N=C_0 \frac{R_2}{R_1}$ the value of the synthetic NC shunt under the assumption of infinite bias resistance $R_0$ \cite{DeManeffePreumont2008,MoheimaniFleming2006}. Other circuit parameters are listed in Tab. \ref{Tab1}.\\
A continuous modulation of $R_2$ allows for a smooth variation of the associated equivalent sandwich stiffness $E_s$, which is function of the shunted Young's modulus $E_{SU}$:
\begin{equation}
E_{s}=\frac{E_{al}I_{al}+2E_p^{SU}I_p}{I_{Al}+2I_p}
\end{equation}
where:
\begin{equation}
I_{al} = \frac{bH^3}{12}, \quad I_p = \frac{bh_p^3}{12}+bh_p(\frac{H}{2}+\frac{h_p}{2})^2
\end{equation}
being $E_{al}=70\cdot10^9$ MPa and $H$ the substrate Young's modulus and thickness. The modulation parameters $\alpha$ and $E_{s,0}$ used Eq. \ref{eq_13} are depending from the maximum and minimum achievable values for $E_{s}\left(t\right)$:
\begin{equation}
\begin{split}
\alpha&=\frac{E_m}{E_{s,0}}=0.275\\[5pt]
E_m&=\frac{E_{s,ON}-E_{s,OFF}}{2}\\[5pt]
E_{s,0}&=\frac{E_{s,ON}+E_{s,OFF}}{2}
\end{split}
\end{equation} 
In a similar way, the equivalent density of the layered part writes:
\begin{equation}
\rho_s= \frac{\rho_{al}A_{al}+2\rho_pA_p}{A_{al}+2A_p}
\end{equation}
which is constant in time, being $\rho_{al} = 2700\;{\rm Kg/m^3}$.
}
\begin{table}[t]
	\begin{tabular}{ c c c c }
		\textbf{Name} & \textbf{Value} & \textbf{Units} & \textbf{Description} \\
		\hline
		$R_1$         & 7.5            & k$\Omega$      &  $-$ \\
		\hline
		$R_2$         & 0-13.7            & k$\Omega$      &  $-$ \\
		\hline
		$R_0$         & 1000           & k$\Omega$      &  \textit{Bias resistance} \\
		\hline
		$C_0$         & 4.4            & nF             &  \textit{NC capacitance} \\
		\hline
		$C_p$         & 7         & nF             &  \textit{piezo patch capacitance} \\
		\hline 
		$d_{31}$      & -1740          & pm/V           &  \textit{piezo strain coefficient} \\
		\hline
		$k_{31}$      & 0.351          & $-$            &  \textit{piezo coupling coefficient} \\	
	\end{tabular}
	\caption{NC shunt circuit parameters. }
	\label{Tab1}
\end{table}
\section{Modal dependece factor}
{Let's consider the $i^{th}$ solution resulting from the eigenvalue problem:
\begin{equation}
({\bm K(\phi)}-\lambda_i {\bm M}) {\bm \Psi}_i = 0 \quad i=1,\dots,n
\label{eq_B1}
\end{equation}
where $\lambda_i=\omega_i^2$ and the eigenvectors are mass normalized, such that ${{\bm \Psi}_i}^T {\bm M} {\bm \Psi}_j = \delta_{ij}$, being $\delta_{ij}$ the Kronecker delta.
Similarly to Fox and Kapoor \cite{fox1968rates}, we compute a sensitivity of $\lambda_i$ with respect to the modulation phase $\phi$, which is representative of the rate of change of $\lambda_i$ in response to a variation of $\phi$.
Differentiating eq. \ref{eq_B1} one gets:
\begin{equation}
\frac{d \lambda_i}{d \phi} = {{\bm\Psi}_i}^T(\frac{d{\bm K(\phi)}}{d\phi}- \lambda_i \frac{d {\bm M}}{d \phi}) {\bm\Psi}_i =  {{\bm\Psi}_i}^T \frac{d{\bm K(\phi)}}{d\phi} {\bm\Psi}_i
\end{equation}
where $dM/d\phi=0$, since the density is not modulated. The eigenvector sensitivity $d\Psi_i/d\phi$ writes:
\begin{equation}
\frac{d {\bm \Psi}_i}{d \phi} = \sum_{r \ne i} \displaystyle\frac{{{\bm \Psi}_i}^T \displaystyle\frac{d{\bm K(\phi)}}{d\phi} {\bm\Psi}_r}{\Delta \lambda_{ir}} {\bm\Psi}_r = \sum_{r \ne i} \displaystyle\frac{\kappa_{ir}}{\Delta \lambda_{ir}} {\bm \Psi}_r
\label{eq_B4}
\end{equation}
where \textit{$\Delta \lambda_{ir}$} is the difference between \textit{$i^{th}$} and \textit{$r^{th}$} eigenvalues and $\kappa_{ir}={{\bm \Psi}_i}^T\frac{d{\bm K(\phi)}}{d\phi} {\bm \Psi}_r$ is the \textit{modal coupling} between \textit{$i^{th}$} and \textit{$r^{th}$} states.
If two eigenvalues $\lambda_i$ and $\lambda_k$ are sufficiently far from the remaining states, such that the term $\Delta\lambda_{ik}$ makes their contribution negligible, the expression \ref{eq_B4} simplifies as:
\begin{equation}
\frac{d {\bm \Psi}_{i,k}}{d \phi} \simeq \frac{\kappa_{ik}}{\Delta \lambda_{ik,ki}} {\bm\Psi}_{k,i}
\label{eq_B5}
\end{equation}
with \textit{$\kappa_{ik} = \kappa_{ki}$}. Now, the effective coupling between states $i$ and $k$ is quantified through the \textit{Modal Dependence Factor} (MDF):
\begin{equation}
MDF_{ik} = \frac{(\frac{\kappa_{ik}}{\Delta \lambda_{ik}})^2}{\sum_{r \ne i} (\frac{\kappa_{ir}}{\Delta \lambda_{ir}})^2}
\end{equation}
which is the ratio between the modal coupling between modes $i,k$ and the coupling of mode $i$ to all modes excepts for itself. A graphical representation of $MDF_{ik}$ is illustrated in \ref{Fig3}(e) $\rm II-III$ for modes 25 and 26 upon varying \textit{$\phi$}.
}

%


\bibliography{References}
\end{document}